\documentclass[lettersize,journal]{IEEEtran}
\usepackage{amsmath,amsfonts}
\usepackage{algorithm}
\usepackage{array}
\usepackage{textcomp}
\usepackage{stfloats}
\usepackage{url}
\usepackage{verbatim}
\usepackage{graphicx}
\usepackage{cite}
\usepackage[english]{babel}
\usepackage{booktabs}

\usepackage{graphicx}
\usepackage{float}
\usepackage{wrapfig}
\usepackage[normalem]{ulem}

\usepackage{bm}
\usepackage{textcomp}
\usepackage{amsmath}
\usepackage{amsthm}
\usepackage{amsfonts}
\usepackage{mathtools}
\usepackage{pifont}

\usepackage{algorithm}
\usepackage{algorithmic}

\usepackage{times}
\usepackage{gensymb}

\usepackage{textcomp}
\usepackage{multirow}
\usepackage{lscape}
\usepackage{subcaption}

\usepackage{mdframed}
\usepackage{hyperref}
\usepackage{textgreek}
\usepackage{tabularx, colortbl, xcolor}
\definecolor{mygray}{gray}{0.95}
\definecolor{mycyan}{HTML}{005397}
\definecolor{myred}{HTML}{E13333}
\definecolor{mymagenta}{HTML}{BF3E87}
\definecolor{mypurple}{HTML}{1B2278}
\definecolor{tearose}{HTML}{F584C5}

\definecolor{coral}{HTML}{F67088}
\definecolor{dodger_blue}{HTML}{3BA3EC}

\definecolor{domino}{HTML}{BC9F48}
\definecolor{catalina_blue}{HTML}{1C3168}
\definecolor{dark_scarlet}{HTML}{C63D52}
\definecolor{cerulean}{HTML}{0192A8}
\definecolor{tussock}{HTML}{C99E31}
\definecolor{p13}{HTML}{BFB5D7}
\definecolor{b14}{HTML}{BEA1A5}
\definecolor{y15}{HTML}{F0Cf61}
\definecolor{Merino}{HTML}{F3EEE3}
\definecolor{Twilight}{HTML}{4E518B}

\newcolumntype{a}{>{\columncolor{p13}}l}

\usepackage[capitalize,noabbrev]{cleveref}
\crefname{ineq}{Inequality}{Inequalities}
\creflabelformat{ineq}{#2{\upshape(#1)}#3} 
\crefname{obj}{Objective}{Objectives}
\creflabelformat{obj}{#2{\upshape(#1)}#3} 
\crefname{exp}{Expression}{Expressions}
\creflabelformat{exp}{#2{\upshape(#1)}#3} 

\theoremstyle{remark}
\newtheorem{example}{Example}[]
\theoremstyle{plain}

\theoremstyle{definition}

\theoremstyle{remark}

\usepackage{braket}
\newcommand{\norm}[1]{\left\lVert#1\right\rVert}
\newcommand{\abs}[1]{\left\lvert#1\right\rvert}

\newcommand{\inner}[1]{\left\langle#1\right\rangle}
\usepackage{stmaryrd}
\newcommand{\iverson}[1]{\left\llbracket#1\right\rrbracket}

\usepackage{tikz}
\usetikzlibrary{trees}
\usepackage{tikz-dependency}
\usepackage{tikz-3dplot}
\usetikzlibrary{chains,scopes}
\usetikzlibrary{arrows.meta,automata,positioning}
\usetikzlibrary{shapes.geometric, arrows}
\usepackage{pgfplots}

\pgfplotsset{
  every axis/.append style = {thick},
  tick style = {thick,black},
  /tikz/normal shift/.code 2 args = {%
    \pgftransformshift{%
        \pgfpointscale{#2}{\pgfplotspointouternormalvectorofticklabelaxis{#1}}%
    }%
  },%
  shift/.style = {
    tick align        = outside,
    scaled ticks      = false,
    enlargelimits     = false,
    ticklabel shift   = {#1},
    axis lines*       = left,
    xtick style       = {normal shift={x}{#1}},
    ytick style       = {normal shift={y}{#1}},
    x axis line style = {normal shift={x}{#1}},
    y axis line style = {normal shift={y}{#1}},
  },
  shift/.default = 10pt,
  shift3d/.style = {
    shift=#1,
    ztick style       = {normal shift={z}{#1}},
    z axis line style = {normal shift={z}{#1}},
  },
  shift3d/.default = 10pt,
}

\tikzstyle{startstop} = [rectangle, rounded corners, minimum width=1cm, minimum height=0.5cm,text centered, draw=black]
\tikzstyle{io} = [trapezium, trapezium left angle=70, trapezium right angle=110, minimum width=2cm, minimum height=0.63cm, text centered, draw=black]
\tikzstyle{process} = [rectangle, minimum width=2cm, minimum height=0.5cm, text centered,draw=black]
\tikzstyle{process_n} = [rectangle, minimum width=1.5cm, minimum height=0.5cm, text centered,text width=1.5cm, draw=black]
\tikzstyle{decision} = [diamond, minimum width=1.2cm, minimum height=0.5cm, text width=1.5cm, text centered, aspect=1.5, draw=black]
\tikzstyle{arrow} = [->,>=stealth]

\usepackage{listings}
\usepackage{array}
\newcolumntype{H}{>{\setbox0=\hbox\bgroup}c<{\egroup}@{}}

\newcommand{\rnumcap}[1]{\MakeUppercase{\romannumeral #1}}

\renewcommand\paragraph[1]{{\vspace{1mm}\noindent\bfseries #1 }}

\usepackage[most]{tcolorbox}
\newtcolorbox[auto counter]{summary}[1][]{title={\bfseries Summary~\thetcbcounter},enhanced,
  coltitle=black,
  colback=white,
  top=0.3in,
  attach boxed title to top left=
  {xshift=1.5em,yshift=-\tcboxedtitleheight/2},boxrule=0.5pt,  sharp corners, fonttitle=\bfseries,boxed title style={size=small,colback=white,colframe=white},#1}

\begin{document}

\title{
A Test Suite for Efficient Robustness Evaluation of Face Recognition Systems
}

\author{Ruihan Zhang, Jun Sun
\thanks{Ruihan Zhang is with Singapore Management University, Singapore. E-mail: rhzhang@smu.edu.sg}
\thanks{Jun Sun is with Singapore Management University, Singapore. E-mail: junsun@smu.edu.sg}}

\markboth{Journal of \LaTeX\ Class Files,~Vol.~14, No.~8, August~2021}%
{Shell \MakeLowercase{\textit{et al.}}: A Sample Article Using IEEEtran.cls for IEEE Journals}


\maketitle

\begin{abstract}
Face recognition is a widely used authentication technology in practice, where robustness is required. It is thus essential to have an efficient and easy-to-use method for evaluating the robustness of (possibly third-party) trained face recognition systems. Existing approaches to evaluating the robustness of face recognition systems are either based on empirical evaluation (\emph{e.g.}, measuring attacking success rate using state-of-the-art attacking methods) or formal analysis (\emph{e.g.}, measuring the Lipschitz constant). While the former demands significant user efforts and expertise, the latter is extremely time-consuming. In pursuit of a comprehensive, efficient, easy-to-use and scalable estimation of the robustness of face recognition systems, we take an old-school alternative approach and introduce \textsc{RobFace}, \emph{i.e.}, evaluation using an optimised test suite. It contains transferable adversarial face images that are designed to comprehensively evaluate a face recognition system's robustness along a variety of dimensions. \textsc{RobFace} is system-agnostic and still consistent with system-specific empirical evaluation or formal analysis. We support this claim through extensive experimental results with various perturbations on multiple face recognition systems. To our knowledge, \textsc{RobFace} is the first system-agnostic robustness estimation test suite.
\end{abstract}


\section{Introduction}
\label{sec:intro}

Face recognition neural networks~\cite{jain2011handbook} are gradually finding applications in security-related tasks such as border control, payment authentication, and automobile access control. Over the past two decades, researchers have developed various face recognition systems with increasing accuracy. However, while state-of-the-art face recognition systems achieve almost perfect accuracy, they are vulnerable to adversarial attacks~\cite{sharif2016accessorize,ramachandra2017presentation,li2020light}. Small perturbations on the input face can lead to completely different predictions from the face recognition system, which is a sign of a lack of robustness. Even if without malicious attacks, lack of robustness results in inconveniences because facial accessories or lighting conditions etc. can prevent users from unlocking their smartphones or prevent monitors in the community from locating lost elderly~\cite{pang2019robust}. When facing malicious attacks, lack of robustness may lead to more severe security risks such as illegal access to other people's bank accounts or failure to identify dangerous criminals~\cite{sharif2016accessorize, nguyen2022master,ramachandra2019custom}. Therefore, before deploying a face recognition system, it is crucial to comprehensively evaluate its robustness under realistic settings. \emph{More importantly, face recognition systems, like many other deep learning systems, may evolve through many versions and each time its robustness must be re-evaluated.} Therefore, we must be able to efficiently evaluate their robustness.

Existing approaches for addressing the problem can be roughly categorized into two groups. The first is to subject the system to extensive attacks and observe its performance~\cite{goswami2018unravelling}. While such an approach is intuitive, it has three shortcomings. First, many attacking methods require constrained searching (\emph{e.g.}, PGD~\cite{madry2017towards}), which is time-consuming if we must evaluate the system's robustness with respect to many kinds of perturbations. Second, proper attacks are also hard to realize in practice because of the variety of attacks~\cite{rauber2020foolbox}. Third, the complex configuration of each attack's setup would often add difficulty for a fair comparison between different (versions of) face recognition systems.

The second is to formally establish certain robustness measures such as Lipschitz constant~\cite{hein2017formal}, \emph{i.e.}, an upper bound on the maximal adversarial effect that can be achieved by any attacker. While these methods provide formal guarantees, they fall short in two aspects. First, it usually only works for a limited measure of robustness. For example, the Lipschitz constant is often used as a theoretical measure of robustness, but only for limited types of systems~\cite{huster2019limitations}. Secondly, it is usually time-consuming. Finding the exact Lipschitz constant for a two-layer neural network is already NP-hard~\cite{virmaux2018lipschitz}, and thus many studies resort to numerical approximations. Although numerical approximation is feasible, it often takes minutes to find a local Lipschitz constant for \textit{a single input sample}~\cite{virmaux2018lipschitz}.

Furthermore, both groups of approaches usually require white-box access to the system, which may not be possible in many settings~\cite{goodman2020advbox}. As a consequence, estimating the robustness of face recognition systems remains a practical problem to be addressed.

\begin{figure}
    \centering
    \includegraphics[width=0.99\linewidth]{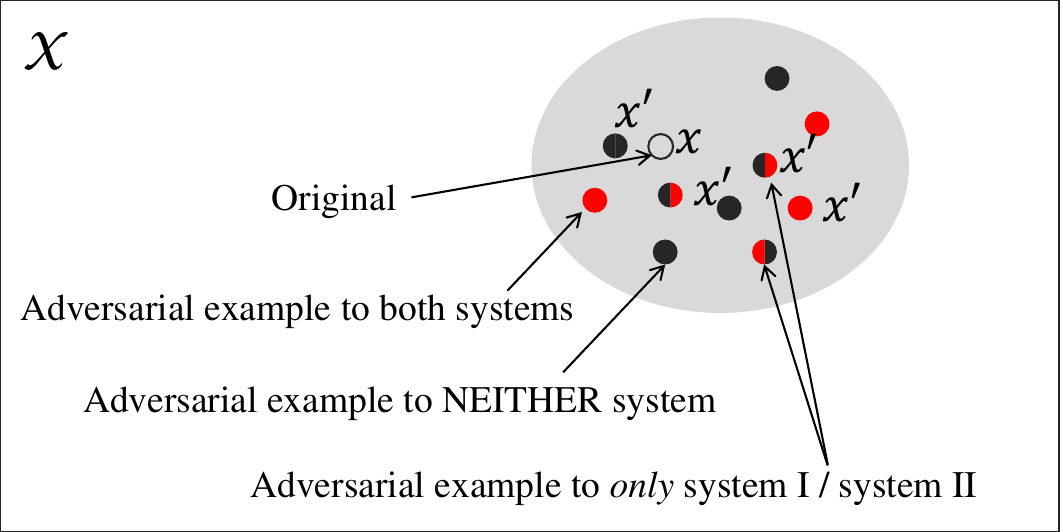}
    \caption{
    Scatter plot of original input $x$ vs perturbed examples $x'$ with colour coding reflecting the effectiveness of the perturbation. Full red dots indicate perturbations that alter predictions for all systems, mixed red and black dots indicate those that alter predictions of only a limited number of systems, and full black dots indicate no effect on altering predictions.
    }
    \label{fig:domain}
\end{figure}

In this work, we propose an old-school yet under-explored approach called \textsc{RobFace}, which leverages an optimised test suite to estimate face recognition system robustness efficiently, \emph{i.e.}, by simply running the system on the tests in the test suite. \textsc{RobFace} is composed of multiple groups of transferable adversarial samples, each of which corresponds to a kind of perturbation, including perturbation under $p$-norms~\cite{goodfellow2014explaining}, facial accessories~\cite{sharif2016accessorize}, and other natural transformations~\cite{zhao2020rdcface}. The aggregated result on each subset is the estimated robustness of the system. 

Compared to the above-mentioned existing approaches, \textsc{RobFace} has three advantages. First, it is convenient to use because the only program that needs to be executed is the face recognition system. Evaluating robustness is carried out in the same way traditional programs are tested. Second, eliminating time-consuming iterative searches in the testing phase significantly speeds up the evaluation process. Third, it provides a comprehensive estimate of robustness, with carefully calibrated test cases in many perturbation dimensions.

The key challenge for a test suite approach like \textsc{RobFace} to work is how to provide accurate robustness evaluation in a system-agnostic way but is still consistent with existing time-consuming approaches such as extensive white-box attacking. \textsc{RobFace} is fundamentally based on the transferability of adversarial perturbation, as illustrated in \cref{fig:domain}. It has been noted in~\cite{demontis2019adversarial,wang2021enhancing,papernot2016transferability} that adversarial perturbation often demonstrates transferability, \emph{i.e.}, the same perturbation may be effective for different inputs or systems. In this work, we take advantage of such a phenomenon to craft an optimised set of adversarial perturbations so that we can evaluate the robustness of unseen systems in a system-agnostic way.

Compared to other test suites/sets, \textsc{RobFace} stresses providing a robustness estimation that aligns well with the theoretical robustness evaluation. This requires the test samples to not only cover a wide perturbation space but also follow an appropriate distribution. Therefore, an optimisation process is necessary to determine the samples in the test suite. In this work, the selection of adversarial samples in each subset is determined by constrained discrete optimisation, where a combination of transferable adversarial examples is shortlisted. This combination can provide estimates that have strong correlations with existing methods. Random face recognition systems are used in the optimisation process, and the validity is supported by testing on other face recognition systems. To prevent ill-minded system developers from inflating the robustness of their system by overfitting to those tests in \textsc{RobFace}, \textsc{RobFace} further provides an option to randomize the test suite based on a secret random seed.

We evaluate \textsc{RobFace} from four perspectives: accuracy, generalizability, efficiency, and diversity. Experiments are conducted by comparing empirical~\cite{madry2017towards} and theoretical~\cite{weng2018evaluating} reference methods. The results show that the estimation of \textsc{RobFace} is consistent with existing empirical and theoretical methods in each perturbation dimension. As for generalizability across different face recognition systems and different perturbations, \textsc{RobFace} is more broadly applicable than existing approaches. Most importantly, \textsc{RobFace} accelerates the robustness evaluation process by more than 200 times over existing approaches.

We highlight the main contributions of this work as follows.
\begin{enumerate}
    \item We propose \textsc{RobFace}, the first search-free and system-agnostic robustness evaluation approach. While providing evaluation results that are consistent with existing approaches, \textsc{RobFace} is 200 times faster.
    \item We construct and publish a pre-optimised test suite \textsc{RobFace}-01, containing adversarial samples corresponding to various perturbations, including $p$-norm and realistic transformations.
    \item We thoroughly examine and report the robustness of existing face recognition systems under each perturbation.
\end{enumerate}

Note that the methodology used to build this test suite can be extended to construct a robustness test suite for tasks other than face recognition. 
In the rest of the manuscript, \cref{sec:preliminary} provides a background overview of face recognition robustness evaluation followed by a clear definition of our problem. \cref{sec:methods} introduces our search-free and system agnostic robustness evaluation of face recognition systems. \cref{sec:experiment}  presents experimental results, and discusses reliability issues and mitigation strategies. \cref{sec:related} discusses related work and \cref{sec:conclusion} is our conclusion.

\paragraph{Data Availability}
All the benchmarks, source code, raw results and statistics are available publicly at \url{https://github.com/cat-claws/RobFace.git}.

\section{Preliminary of Face Recognition and Robustness Testing}
\label{sec:preliminary}

In the following, we provide the necessary preliminaries for studying this topic, including notations, formalisation of existing approaches, and challenges. At the end of this section, we define our problem.

\subsection{Background}

\begin{figure}[t]
    \centering
    \begin{subfigure}{0.32\linewidth}
	\centering
	\includegraphics[width=\linewidth]{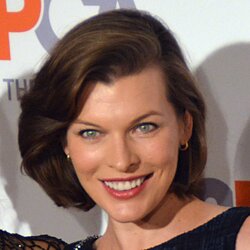}
	\caption{}
    \end{subfigure}
    \begin{subfigure}{0.32\linewidth}
	\centering
	\includegraphics[width=\linewidth]{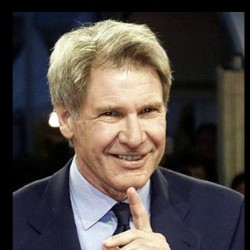}
	\caption{}
    \end{subfigure}
    \begin{subfigure}{0.32\linewidth}
	\centering
	\includegraphics[width=\linewidth]{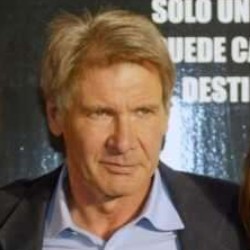}
	\caption{}
    \end{subfigure}
    \caption{Displayed above are three face images, each with a resolution of 250$\times$250 pixels. In face recognition, an input could look like $\bm{x}_1 = (\textnormal{im}_\textnormal{a}, \textnormal{im}_\textnormal{b})$ or $\bm{x}_2 = (\textnormal{im}_\textnormal{b}, \textnormal{im}_\textnormal{c})$. Here, $n=6.25\time 10^4$. Correct predictions of a classifier $h$ should be $h(\bm{x}_1) = 0$ (different persons) and $h(\bm{x}_2) = 1$ (same person).
    \label{fig:facereco}
    }
\end{figure}
\paragraph{Face Recognition and Face Encoding.}
The face recognition task involves predicting if two face images belong to the same individual~\cite{jain2011handbook}. In this binary classification task, a classifier $h$ receives \emph{an} input that represents \emph{both} face images. Suppose each image comes from $\mathbb{R}^n$, the input to $h$ can be seen as an ordered pair of two $n$-dimensional vectors, \emph{i.e.}, $\bm{x} = (\bm{x}^\alpha, \bm{x}^\beta)$ and $\bm{x}\in\mathbb{R}^n\times\mathbb{R}^n$. Thus, $h$ can be formally expressed as
\begin{equation}
\label{eq:h}
    h: \mathbb{R}^n\times\mathbb{R}^n \to \{0, 1\}.
\end{equation}

In many practical cases, the function $h$ is decomposed and we can find a function $g$ such that
\begin{equation}
\label{eq:decompose}
    h(\bm{x}) = \mathbf{1}_{\inner{g(\bm{x}^\alpha), g(\bm{x}^\beta)} > \kappa}
\end{equation}
where $\mathbf{1}_{\textnormal{condition}}$ is returns 1 if and only if the condition is true~\cite{iverson1962programming,bracewell1986fourier}, $\inner{\cdot}$ denotes inner product, and $\kappa$ is some threshold value in $(-1, 1)$. This kind of decomposable $h$ is typically referred to as Siamese system~\cite{chicco2021siamese} where $g$ is its face feature extractor/encoder.


The state-of-the-art face encoders are usually neural networks, such as convolutional neural networks~\cite{wu2018light,jain2011handbook} or vision transformers~\cite{xia2022vision}. They encode each face image into a dense vector and the inner product of the two encoded vectors measures their closeness. If the inner product exceeds a threshold, the two faces are predicted to belong to the same person.

\paragraph{Robustness Testing.}
The robustness of a system quantifies how unlikely the system's prediction changes when input is perturbed. Formally, the worst-case robustness~\cite{li2023sok,pang2022robustness} of a classification system $h:\mathbb{X}\to\mathbb{Y}$ can be expressed as
\begin{equation}
\label{eq:robustness}
    P_{(\mathbf{x}, \textnormal{y})\sim D} \Big(\lnot \exists \bm{x'} \in \mathbb{X}.\quad  (d(\mathbf{x},\bm{x'}) < \epsilon) \land (h(\bm{x'}) \neq \textnormal{y}) \Big)
\end{equation}
where $D$ is the joint distribution over $\mathbb{X}\times\mathbb{Y}$, and $(\mathbf{x},\textnormal{y})$ is its random variable.

$d$ is a distance function and $\epsilon$ is a threshold. Intuitively, any point whose distance to a given input centre is lower than a threshold is considered within a vicinity~\cite{ma2018characterizing}, or simply a neighbour of the centre. The distance function $d$ can take various forms,
such as the Euclidean distance between two vectors (in $\mathbb{X}$). This can be generalised into an $L^p$-norm where $p$ can be $0, 1, 2, \ldots, \infty$. Furthermore, natural transformation~\cite{bhattacharya2019survey} such as radial distortion can also be captured by $d$ using a piecewise function, \emph{i.e.}, distance is 0 when a point can be obtained from the natural transformation of the other, and $\epsilon + 1$ otherwise.



There could be alternative or more generalised definition of robustness~\cite{zhang2024certified,goodfellow2014explaining}, which actually is equivalent to \cref{eq:robustness} when it comes to classification problems. \cref{eq:robustness} does not restrict the input format. An input can be a vector~\cite{wiyatno2019adversarial} (of tabular data), an image~\cite{goodfellow2014explaining}, or a language unit~\cite{yoo-etal-2020-searching}. For face recognition, input $\bm{x}$ is an ordered pair of images as shown in \cref{fig:facereco}, and $\mathbb{X}\subset \mathbb{R}^n\times\mathbb{R}^n, \mathbb{Y}=\set{0,1}$.

Although \cref{eq:robustness} provides a method for calculating the robustness of a given system $h$ on a distribution $D$, implementing an algorithm to compute this value remains challenging. Sampling $(\mathbf{x}, \textnormal{y})$ from $D$ is straightforward (a finite dataset $\mathbb{S}$ often serves as sampling from the distribution $D$, and the sample mean often approximates the expectation), but verifying if all neighbours $\bm{x'}$ of the original input satisfy $h(\bm{x'})=\textnormal{y}$ is difficult.

Practical evaluation approaches fall into two categories: 1) empirical testing, which involves time-consuming iterative searches for adversarial examples, \emph{i.e.}, some neighbour $\bm{x'}$ satisfying $h(\bm{x'})\neq\textnormal{y}$, and 2) formal verification, which fully guarantees the absence of adversarial examples but often remains incomplete as sometimes a `not sure' is thrown. This study aims to increase the efficiency and user-friendliness of empirical testing.

\paragraph{Robustness Testing of Face Recognition system.}
Perturbing the input to a face recognition system may change its prediction.
Since input $\bm{x}$ of a system contains two face images, the perturbation may occur on either face or both. In common settings~\cite{dong2019efficient}, only one face ($\bm{x}^\alpha$) is perturbed.



Robustness issues in face recognition can arise from various factors, such as camera noise or camera angle. Vulnerability to one perturbation scheme does not necessarily imply vulnerability to another. Therefore, robustness must be tested for respective vicinity $d$ and $\epsilon$ to ensure comprehensiveness~\cite{huster2019limitations}. This work aims to comprehensively estimate the robustness of a face recognition system $h$.

\begin{example}
Assume that we would like to know how robust a face recognition system is with respect to two types of perturbations, \emph{i.e.}, wearing glasses and radial distortion. To evaluate this, we first set limits on how much glasses can cover the face (\emph{e.g.}, no more than 5\% or 10\% of the image) and how curved the face can be (\emph{e.g.}, no more than 10 degrees or 15 degrees). We can then test the system's robustness and report the results in the form of four scalar values.
\end{example}

\subsection{Existing System-in-the-loop Robustness Testing}
\label{sec:adversarial}
\begin{figure*}[t]
    \centering
    \subcaptionbox{Ours\label{fig:overview-ours}}{%
      \includegraphics[height=4cm]{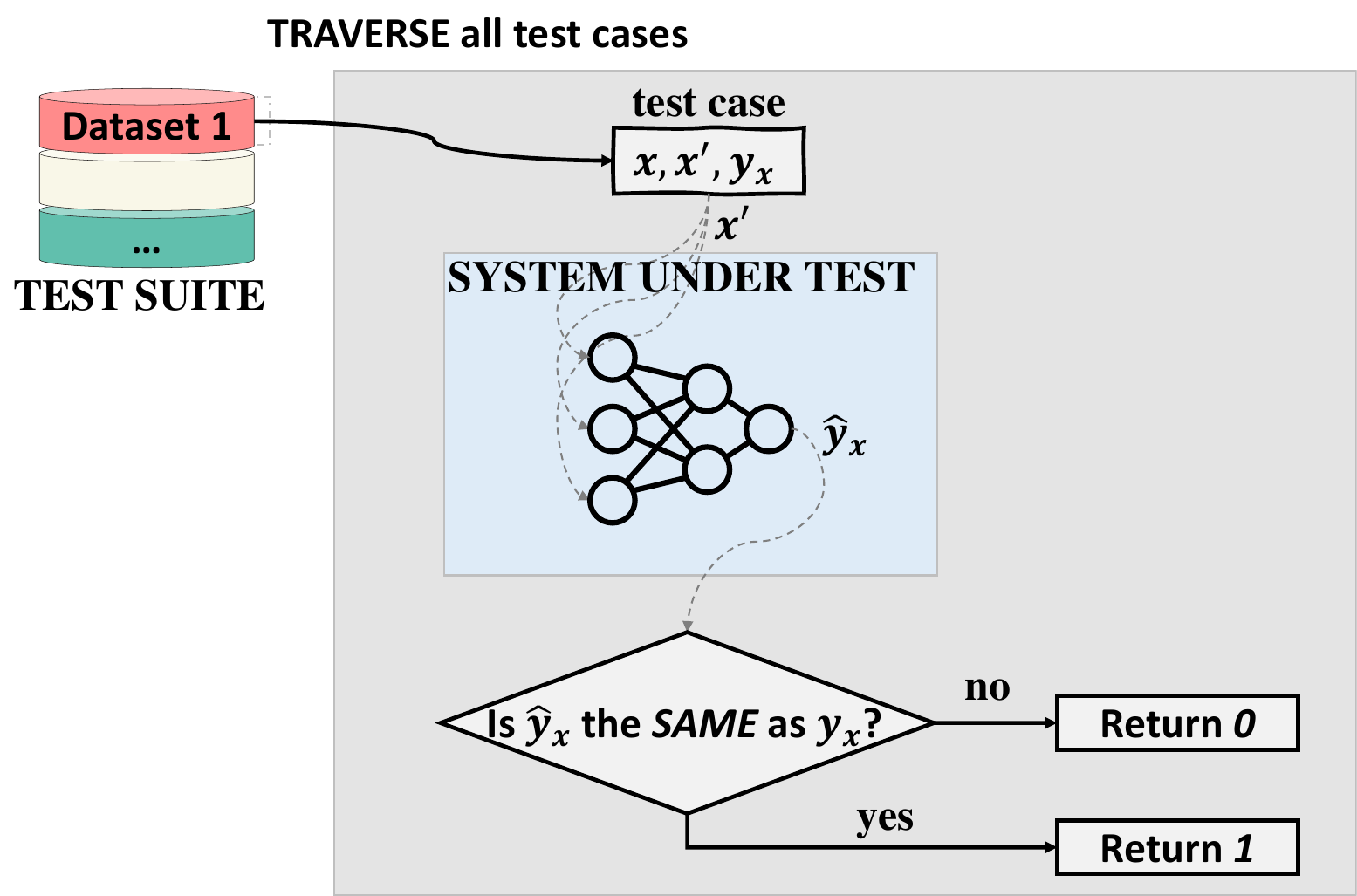}%
    }\quad\quad\quad\quad\quad\quad
    \subcaptionbox{Empirical evaluation\label{fig:overview-empirical}}{%
      \includegraphics[height=4cm]{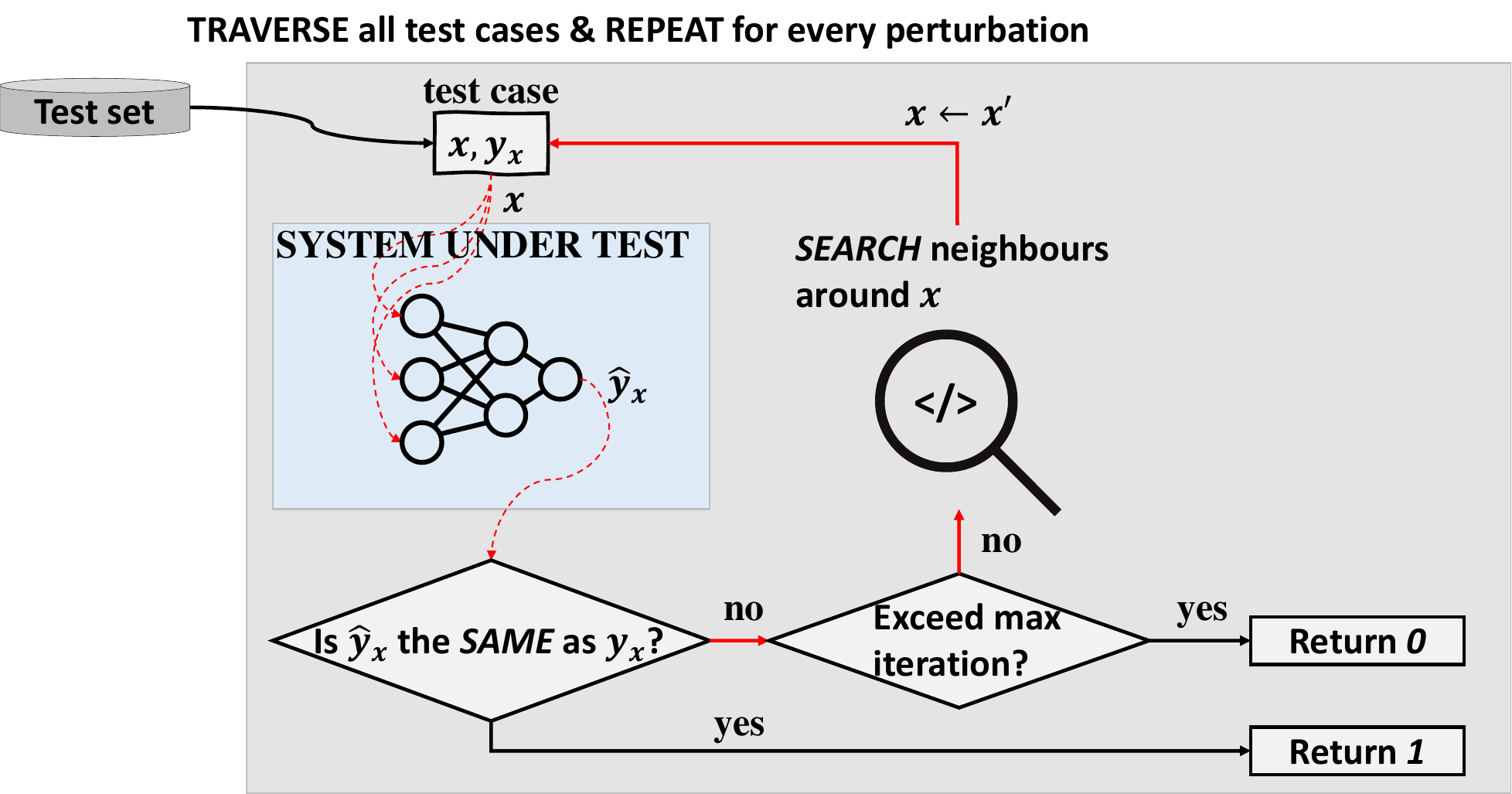}%
    }   
    \caption{Differences in robustness estimating space. The left shows the process of evaluating robustness using a test suite (\emph{i.e.}, what we propose in this work). The right shows the process of empirical evaluation using adversarial attacks, which require us to iteratively approximate the maximum change in the output through optimisation (shown as a loop).}
    \label{fig:overview}
\end{figure*}

The exact computation of \cref{eq:robustness} is computationally intractable~\cite{carlini2019evaluating}, and multiple methods have been proposed to numerically approximate it through adversarial attacks or Lipschitz regularity.

\paragraph{Mechanism of Adversarial Attacks.}
An adversarial attack is a local search algorithm that iteratively explores the neighbours $\bm{x'}$ of a centre input $\bm{x}$. The iteration stops if at least one adversarial example is detected, \emph{i.e.}, $h(\bm{x'}) \neq y$, or if the maximum number of iterations $s_{\max}$ is reached~\cite{goodfellow2014explaining, kurakin2016adversarial}.

To estimate the likelihood of $h(\bm{x'}) \neq y$, a loss function $l(h, \bm{x'}, y)$ is used where a higher loss means it is more likely that $h(\bm{x'}) \neq y$. Identifying the max-loss neighbour around an input $\bm{x}$ is usually formulated as an optimisation problem, which is solved differently depending on whether $h$ is a white box or a black box.

In a white-box setting, the partial gradient $\nabla_{\bm{x'}} l(h, \bm{x'}, y)$ can be numerically obtained with respect to any perturbed sample $\bm{x'}\in\mathbb{X}$. With the gradient, the local maximum can be approximated through gradient descent algorithms. The PGD attack~\cite{madry2017towards} is a widely used white-box method~\cite{croce2020scaling}. It iteratively perturbs the input $\bm{x}$ in the direction of the gradient of the loss function $l(h, \bm{x'}, y)$ with respect to the input~\cite{madry2017towards}. At each iteration step $s_t$, the perturbation is projected back onto the vicinity around $\bm{x}$. Formally,
\begin{equation}
\label{eq:pgd}
    \bm{x'}_{s_{(t+1)}} = \epsilon\frac{\bm{x'}_{s_{(t)}} + \eta_a \cdot \text{sign}(\nabla_{\bm{x'}_{s_{(t)}}} l(h, \bm{x'}_{s_{(t)}}, y))}{\norm{\bm{x'}_{s_{(t)}} + \eta_a \cdot \text{sign}(\nabla_{\bm{x'}_{s_{(t)}}} l(h, \bm{x'}_{s_{(t)}}, y))}}
\end{equation}
where $\eta_a$ is the step size and the norm on the denominator projects onto the border of the vicinity.

White-box tests in face recognition require the systems to provide the confidence level $\inner{g(\bm{x'}^\alpha), g(\bm{x'}^\beta)}$, and gradients $\nabla_{\bm{x'}} \inner{g(\bm{x'}^\alpha), g(\bm{x'}^\beta)}$ that indicate the fastest direction to change the confidence level~\cite{szegedy2013intriguing,carlini2017towards}. White-box attacks are generally effective due to this information. As shown in \cref{fig:overview-empirical}, the system $h$ is called by the attack algorithm at each step, and practically, the number of steps ranges from 20~\cite{robey2022probabilistically} to 200~\cite{croce2020scaling}. Moreover, the white-box processes (red lines in \cref{fig:overview-empirical}) preserve gradients of $h$, and thus halves its forward prediction speed.

In black-box mode, the system provides predictions for the current neighbours $\set{\bm{x'}, \ldots}_{s_t}$, and the attack algorithm compares these predictions and centre label $y$ to guide the search~\cite{papernot2017practical,brendel2017decision}.

In a black-box setting, we do not know the parameters of $l(h, \bm{x'}, y)$. Therefore, the derivative is not available even by numerical computation. In such cases, differential evolution~\cite{su2019one} or genetic algorithms~\cite{alzantot2019genattack} are generally adopted. Formally, for the finite set $\hat{B}(\bm{x}) \subset \set{t \mid t\in\mathbb{X}, d(\bm{x},t) < \epsilon}$ that contains all iterated neighbours of $\bm{x}$ in the sampled set $\mathbb{\hat{X}}$ during optimisation, robustness $r_\gamma(h)$ is approximated as follows, where $\gamma$ denotes a specific perturbation scheme, e.g., 8/255 in $L^\infty$-norm,
\begin{equation}
    \begin{aligned}
    r_\gamma&(h)\\
    &=\underset{\mathbb{\hat{X}}}{\operatorname{Mean}} \left(\iverson{\bigg(\min_{t \in \hat{B}(\bm{x}), y' \neq y} l(h, t, y') - l(h, t, y)\bigg) > 0}\right)  
    \end{aligned}
\end{equation}
The overall process of empirical robustness evaluation is shown on the right of~\cref{fig:overview}. For comparison, the process of our approach is shown on the left.

Although white-box attacks can be strong in searching adversarial examples, they typically require more information from systems compared with black-box attacks, which in turn limits the application of white-box attacks~\cite{papernot2017practical,brendel2017decision}.



\paragraph{Robustness Verification with Lipschitz Regularity.}
This category of methods does not iteratively search for adversarial examples but still needs to repeatedly call the system many times for each input sample~\cite{weng2018evaluating}. It verifies the system as a white box, such as by studying the Lipschitz regularity of the neural model's last layer output~\cite{weng2018evaluating}. The verification methods are theoretically developed for estimating true robustness, and results usually correlate with those from attacks.

A representative method is to calculate the local Lipschitz constant $K_q$ of the real-valued function $f$ such that
\begin{equation}\label[ineq]{ineq:continuous}
	\abs{f(\bm{x}_1) - f(\bm{x}_2)} \leq K_q \norm{\bm{x}_1 - \bm{x}_2}_p, \forall \bm{x}_1, \bm{x}_2 \in \mathbb{R}^n
\end{equation}
where $p^{-1} + q^{-1} = 1$ and $1\leq p, q \leq\infty$. \emph{The} Lipschitz constant is defined as the smallest $K_q$ that satisfies \cref{ineq:continuous}. As such, $K_q$ serves as the maximum norm of a function's gradient across its input domain~\cite{weng2018evaluating}, as follows.
\begin{equation}\label{eq:lipschitz}
    K_q(f) = \sup_{\bm{x}\in\mathbb{R}^n} \norm{\nabla f(\bm{x})}_q
\end{equation}
When it comes to measuring the robustness, the domain is as described in the vicinity of an input. Therefore the Lipschitz constant of an input $\bm{x}$ caps the maximum loss in its neighbourhood, expressed as follows.
\begin{equation}\label{eq:lipschitz_2}
    K_q(f) = \sup_{\bm{x'}\in\mathbb{X}, d(\bm{x},\bm{x'}) < \epsilon} \norm{\nabla h(\bm{x})}_q
\end{equation}
Although Lipschitz regularity theoretically guarantees a lower bound of robustness~\cite{weng2018evaluating}, computing the Lipschitz constant is known to be NP-hard if the system is a two-layer neural network or more complex~\cite{virmaux2018lipschitz,hein2017formal}. Thus, this method is computationally intractable in practice. Moreover, its theory is only applicable if the transformation in the input space ($d$) is measured in $p$-norms. Many transformations from a sample to a perturbed sample that cannot be characterised by $p$-norm are relevant in practice, such as brightness, spatial offset or scaling in the image. When facing such real-world transformation other than $p$-norm, the theory must be extended (and we do not know how yet).

There is an alternative line of research on formal verification of neural networks~\cite{singh2019abstract,katz2017reluplex}. Such approaches tend to return ``unknown'' as an answer or fall short in evaluating large systems like the recent face recognition systems. Therefore, these approaches are not adopted as references in this work.

\subsection{Problem definition}

Our goal is to develop an efficient and system-agnostic method for empirical robustness testing in face recognition. The challenge is to enhance efficiency without compromising the accuracy of robustness testing. Additionally, we aim to comprehensively address various perturbations.

Formally, when evaluating the robustness of a given face recognition system $h$, we need to consider various different perturbations $(d_1, \epsilon_1), (d_2, \epsilon_2), \ldots (d_\gamma, \epsilon_\gamma), \ldots$. For each perturbation scheme $\gamma$, e.g., an $L^2$ norm with a size of 7, we would be given a reference robustness value $r_\gamma$, obtained from state-of-the-art, typically time-consuming robustness testing algorithms, such as PGD~\cite{madry2017towards}. In the following, we denote our predicted robustness as $\Tilde{r}_\gamma$, and outline the proposed method's requirements.

\begin{enumerate}
    \item A high correlation between our evaluation and the reference robustness testing values is required. Given some system (as a random variable) $\mathbf{h}$ from the system space $\set{\mathbb{X}\to\{0,1\}}$, the correlation between reference robustness $\textnormal{r}_\gamma$ and our estimated robustness $\Tilde{\textnormal{r}}_\gamma$, as captured by \cref{eq:corr}, should be statistically significant.
    \begin{equation}
    \label{eq:corr}
        \frac{\operatorname{E} \left[(\Tilde{\textnormal{r}}_\gamma-\mu _{\Tilde{\textnormal{r}}_\gamma})(\textnormal{r}_\gamma-\mu _{\textnormal{r}_\gamma})\right]}{\sigma _{\Tilde{\textnormal{r}}_\gamma}\sigma _{\textnormal{r}_\gamma}}
    \end{equation}
    \item This correlation needs to be generalisable on unseen face recognition systems.
    \item Our robustness testing should take less time and computation resources in runtime, compared with obtaining the reference.
    \item Robustness testing of face recognition systems should comprehensively consider various perturbation schemes. Apart from $p$-norm constrained perturbations, these schemes also include a variety of practical transformation-based perturbations~\cite{zhao2020rdcface}.
\end{enumerate}

\section{Search-free and System-agnostic Robustness Evaluation for Face Recognition}
\label{sec:methods}
We propose a search-free system-agnostic approach for evaluating the robustness of face recognition systems. That is, we build a suite of well-crafted test cases, which can be used to efficiently assess the robustness of these systems. The approach is novel and the usage of the test suite is easy to understand, illustrated on the left of~\cref{fig:overview}. In the following, we first explain how to use the test suite, and then describe how it is built to satisfy the above-mentioned requirements.

Mathematically, the test suite is a set of sets, and we 
denote it as follows.
\begin{equation}
\label{eq:testsuite}
    \mathcal{T} \coloneqq \set{\mathbb{T}_1,\ldots, \mathbb{T}_\gamma, \ldots}
\end{equation}
As shown, the test suite has multiple categories ($\mathbb{T}$) according to different perturbations $d, \epsilon$. Note that some perturbation schemes may have the same $d$ but different $\epsilon$. For any perturbation, $\mathbb{T}_\gamma$ is a parallel test set where each element is an ordered triple $(\bm{x}, \bm{x'}, y)$.
In the ordered triple, we let the first entry be the original input (pair of two original faces), and the second entry be a \textbf{\emph{carefully}} perturbed input ($\bm{x}^\alpha$ has been perturbed). The third entry is the centre label $y$.

Intuitively, the size of $\mathcal{T}$ suggests how many kinds of perturbations are considered, i.e., how many distinct $\gamma$. The size of each $\mathbb{T}$ suggests how many (triples of) test cases are there in this category. The number of test cases in each category may vary. From a practical data format perspective, the set can be optimised by saving one original vector per source and avoiding duplicating unmodified faces.

We present our method in two parts. We first show how to use such a test suite to evaluate robustness. Second, we present how might we construct such a demanded test suite.

\subsection{Evaluating Robustness of Face Recognition System Using a Test Suite}

In the following, we show the runtime usage of our method, \emph{i.e.}, how to use a test suite to evaluate robustness. Using a test suite to evaluate robustness is as simple as using a test set to evaluate the system's accuracy. During inference, it does not require any optimisation, iterative search, or repeated running system regarding one sample, which significantly boosts efficiency.

Given a test suite $\mathcal{T}$, we use each $\mathbb{T}_\gamma$ in the following way. First, for each $(\bm{x}, \bm{x'})$ in the triple, a system makes a prediction about whether $ h(\bm{x}) = h(\bm{x'})$. Note that $y$ in the triple is for reference only. Next, we take the number of $ h(\bm{x}) = h(\bm{x'})$ cases divided by the total number as our estimated robustness $\Tilde{{r}}_\gamma(h)$. Formally,
\begin{equation}\label{eq:evaluate}
    \Tilde{{r}}_\gamma(h) \coloneqq 1 - P_{(\bm{x}, \bm{x'}, y)\sim \mathcal{U}(\mathbb{T}_\gamma)} \big( h(x') \neq  y\big)
\end{equation}

This process applies robustness evaluation for a variety of perturbation schemes. Thus, we can follow the same procedure for the remaining sets in the test suite and obtain a comprehensive robustness evaluation of the face recognition system under test. We name the robustness evaluation using $\mathcal{T}$ as \textsc{RobFace}. An intuitive illustration is given in \cref{fig:overview-ours}, as compared against a more complicated procedure of existing methods shown in \cref{fig:overview-empirical}.

\begin{example}
Assume that we would like to evaluate the robustness of ID-DNN~\cite{wang2017unleash} under a glass perturbation (\emph{i.e.}, the perturbation adds a pair of glass on the images). To do that, we select the corresponding dataset from our prepared test suite. It contains 2,334 samples, where the glasses cover no more than 5\% of each image. Each sample consists of two 37,632-dimensional (3$\times$112$\times$112) vectors concatenated together to form a 75,264-dimensional vector. We input each sample into ID-DNN, which outputs a binary prediction of either 0 or 1. If the prediction matches the label for the original sample, we count it as true, otherwise false. We calculate the percentage of true predictions to measure the model's robustness under this specific perturbation.
\end{example}

\begin{figure}[t]

    \begin{mdframed}[backgroundcolor=Merino!99,linecolor=Merino!99]
    \centering
    \begin{subfigure}{0.235\linewidth}
	\centering
	\includegraphics[width=0.98\linewidth]{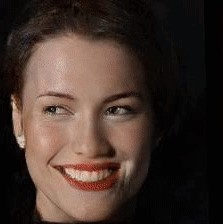}
	\caption*{\scriptsize $L^2 \leq 7$}
	\end{subfigure}
	\begin{subfigure}{0.235\linewidth}
	\centering
	\includegraphics[width=0.98\linewidth]{figure/n1.png}
	\caption*{\scriptsize $L^2 \leq 10$}
	\end{subfigure}
	\begin{subfigure}{0.235\linewidth}
	\centering
	\includegraphics[width=0.98\linewidth]{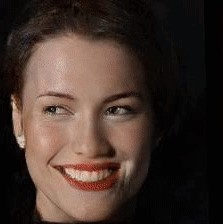}
	\caption*{\scriptsize $L^\infty \leq 1$}
	\end{subfigure}
	\begin{subfigure}{0.235\linewidth}
	\centering
	\includegraphics[width=0.98\linewidth]{figure/n2.png}
	\caption*{\scriptsize $L^\infty \leq 4$}
	\end{subfigure}\\\vspace{1mm}
    \begin{subfigure}{0.235\linewidth}
	\centering
	\includegraphics[width=0.98\linewidth]{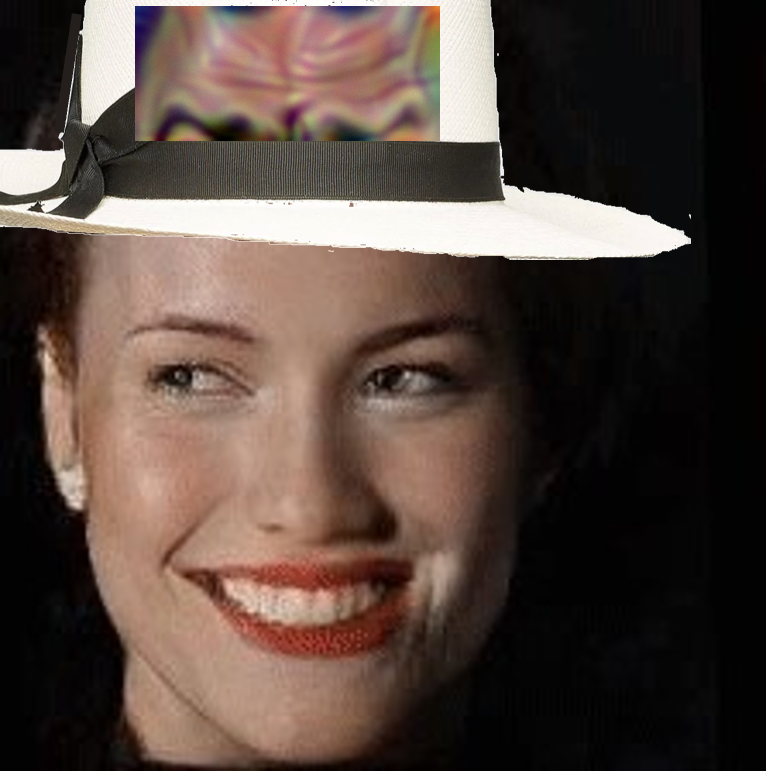}
	\caption*{\scriptsize Patch \tiny $\leq 10\%$ ~\cite{komkov2021advhat}}
	\end{subfigure}
	\begin{subfigure}{0.235\linewidth}
	\centering
	\includegraphics[width=0.98\linewidth]{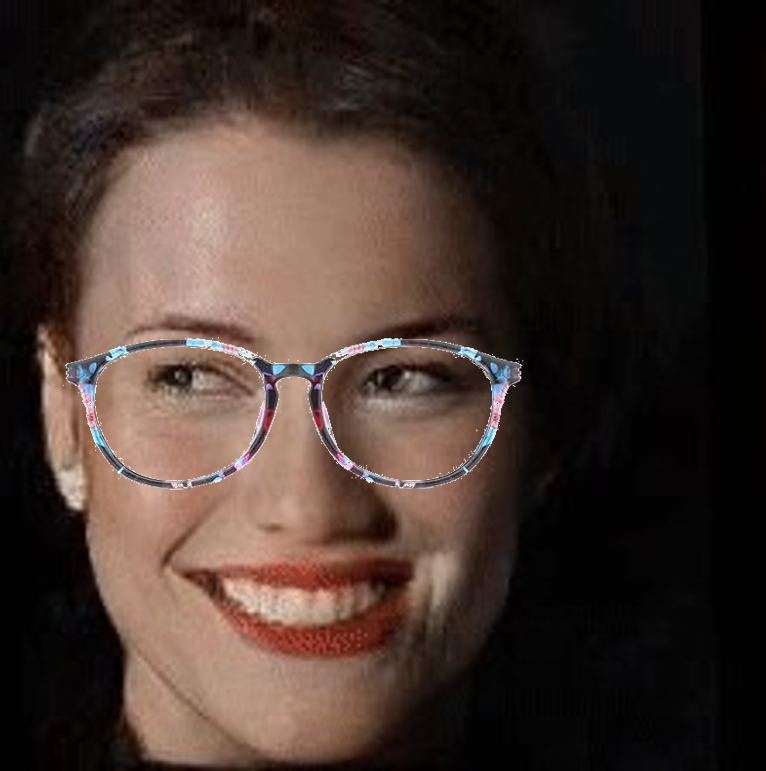}
	\caption*{\scriptsize Glasses $\leq7\%$~\cite{sharif2016accessorize}}
	\end{subfigure}
	\begin{subfigure}{0.235\linewidth}
	\centering
	\includegraphics[width=0.98\linewidth]{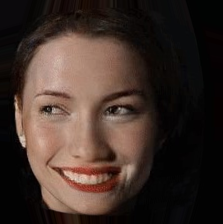}
	\caption*{\scriptsize Radial $\leq84\degree$~\cite{zhao2020rdcface}}
	\end{subfigure}
	\begin{subfigure}{0.235\linewidth}
	\centering
	\includegraphics[width=0.98\linewidth]{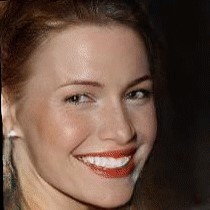}
	\caption*{\scriptsize Pose $\leq45\degree$~\cite{zheng2018cross}}
	\end{subfigure}\\\vspace{1mm}
	\begin{subfigure}{0.235\linewidth}
	\centering
	\includegraphics[width=0.98\linewidth]{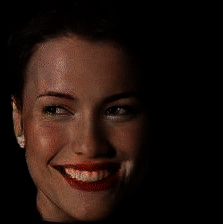}
	\caption*{\scriptsize Illumination~\cite{zou2007illumination}}
	\end{subfigure}
    \begin{subfigure}{0.235\linewidth}
	\centering
	\includegraphics[width=0.98\linewidth]{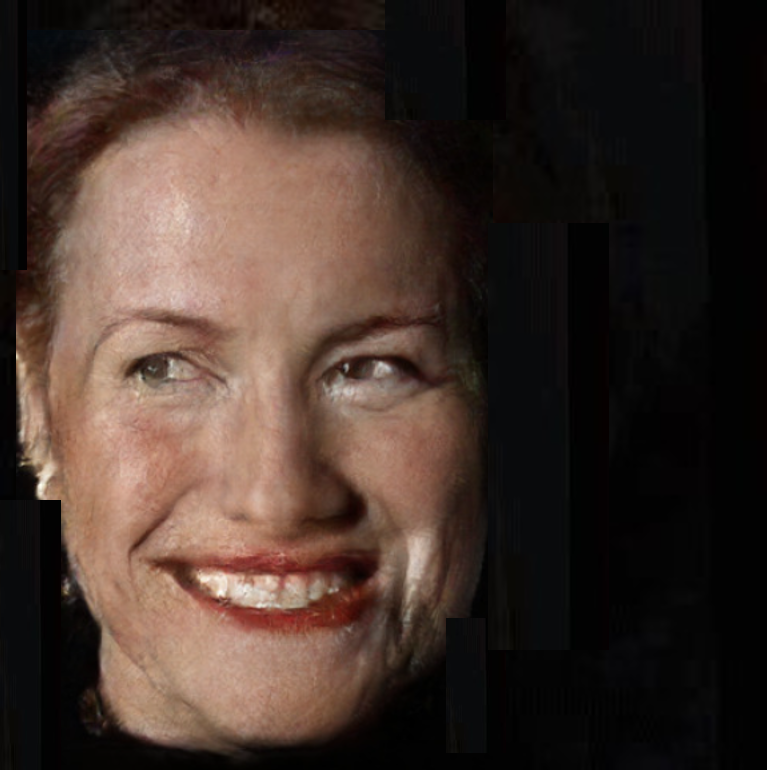}
	\caption*{\scriptsize Age change~\cite{zheng2018cross}}
	\end{subfigure}
	\begin{subfigure}{0.235\linewidth}
	\centering
	\includegraphics[width=0.98\linewidth]{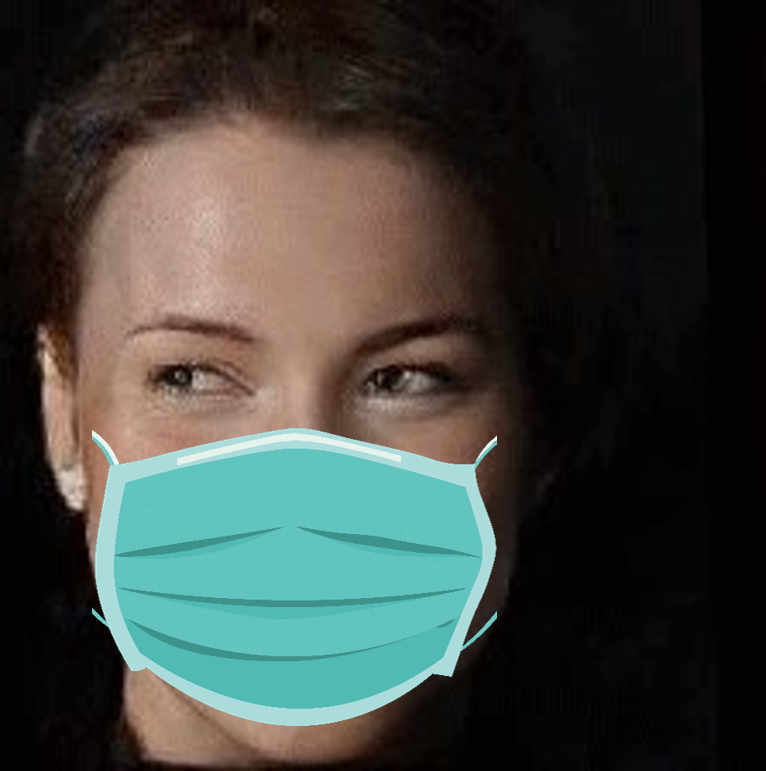}
	\caption*{\scriptsize Face mask~\cite{wang2022mlfw}}
	\end{subfigure}
	\begin{subfigure}{0.235\linewidth}
	\centering
	\includegraphics[width=0.98\linewidth]{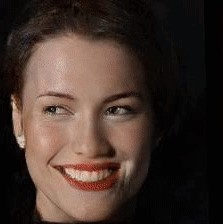}
	\caption*{\scriptsize Original}
	\end{subfigure}\\
    \end{mdframed}
    \caption{Illustration of various face image perturbations, including perturbation within certain $p-$norm bounds and other perturbations such as wearing unique accessories, altering ambient light or background, and adjusting camera angle. Some transformations have additional restrictions such as illumination intensity and age range.
    }
    \label{fig:perturbations}
\end{figure}

\subsection{Constructing a High-quality Test Suite for Evaluating Robustness of Face Recognition}
\label{sec:construct}
While using the test suite is straightforward as we show above, building the test suite $\mathcal{T}$ is highly nontrivial. There are three main steps as we explain below. First, we examine a variety of the perturbation space defined as ($d, \epsilon$) to cover as many relevant perturbations as possible. Next, for each perturbation space, we generate faces that can potentially serve as adversarial examples for multiple face recognition systems. Finally, we select from the generated examples such that when different face recognition systems are tested, their robustness is evaluated accurately and efficiently.

\paragraph{Determining Perturbations in Test Suite.}~~ In the real world, a face recognition system must be robust to various types of perturbations. Therefore, we must examine different perturbation spaces $(d, \epsilon)$ for a comprehensive evaluation of face recognition systems. In this step, we identify a comprehensive collection of relevant face perturbations from the literature. Note that different perturbations may share the same $d$ but have different $\epsilon$, \emph{e.g.}, a person can wear glasses of different sizes. In this work, we consider the perturbation spaces shown in \cref{fig:perturbations}, including $p$-norm and a variety of transformations to faces. We remark that the list of perturbations can easily be expanded if other perturbations become relevant in the future. \\

\begin{algorithm}[tb]
\caption{Candidate Adversary Generation \textsc{(\textsc{RobFace-GEN})} }\label{alg:gen}
\begin{algorithmic}[1]
\STATE {\bfseries Input:} Finite parallel data set $\mathbb{S}$, distance function $d$, distance upper bound $\epsilon$,  dummy face \textit{encoding} system $g^\dagger$.
\STATE {\bfseries Hyperparameter:} Number of steps $N$, step size $\eta$
\STATE {\bfseries Initialization:} Empty set $\mathbb{T'}_\gamma$
\STATE {\bfseries Output:} Populated set $\mathbb{T'}_\gamma$
\FORALL {$(\bm{x}, \bm{x'}, y) \in \mathbb{S}$}
\STATE $l \gets 1, ~~\epsilon'\gets\operatorname{random}(0, \epsilon), ~~i\gets 0$
\WHILE{$i < N, ~~i \gets i + 1$}
\STATE $x'\gets d^{-1}_{\epsilon'}(x)$
\STATE $t_\textsc{Inner} \gets \inner{~g^\dagger\big(\bm{x'}^\alpha\big),~g^\dagger\big(\bm{x'}^\beta\big)~}$
\STATE $l' \gets 2  y  t_\textsc{Inner} - t_\textsc{Inner}$
\STATE $\epsilon' \gets \epsilon' - \eta\nabla(l')$
\STATE $x' \gets \operatorname{Clip}\big(x', d^{-1}_{\epsilon}(x)\big)$ 
\IF{$l' < l$}
\STATE $\mathbb{T'}_\gamma \gets \mathbb{T'}_\gamma\cup\set{(\bm{x}, \bm{x'})}$
\STATE $l \gets l'$
\ENDIF
\ENDWHILE
\ENDFOR
\end{algorithmic}
\end{algorithm}

\paragraph{Generating Potentially Plausible Adversarial Examples.}
The outcome of this step is a plausible adversary set $\mathbb{T'}_\gamma$ that contains a set of paired vectors. Each pair consists of an original input vector from the standard test set (\emph{e.g.}, LFW~\cite{huang2008labeled}) and its corresponding perturbed input vector. Retaining the original vector pairs allows us to measure the distance and determine the label. The original input vectors are modified in a controlled and gradual manner, using the gradient descent algorithm, until a certain number of steps have been taken. The perturbations applied to each input are constrained by a distance function and a distance upper bound.

Particularly, \cref{alg:gen}  (\textsc{RobFace}-GEN) \textit{gradually} adds pairs of vectors to an empty set, illustrated in \cref{alg:gen}. We start by initializing an empty set $\mathbb{T'}_\gamma$. Then, for each input in the given test set, it iteratively generates perturbed vectors. Throughout the iteration, the process involves perturbing this input in a monotonic manner, meaning that a new perturbation is only added to the set if it results in a greater prediction difference compared to any previous perturbation. Consequently, our plausible adversary set represents a spread of prediction differences from very similar to very different from the original input. As a result, the generated perturbed samples are close to the original in the \textit{input} space, but have a range of differences in the prediction space, from barely noticeable to highly noticeable. This can help increase the diversity of sample features and make them potential adversarial examples for other face recognition systems.

Another benefit of \textsc{RobFace-GEN} is that it can manipulate transformation parameters, such as illumination and radial angle. This allows us to generate a large number of perturbed samples under each perturbation scheme, which is difficult to achieve with other methods like universal adversarial example search~\cite{demontis2019adversarial}. Additionally, generating a high number of samples does not result in increased computational expenses or longer processing times. 


\begin{example}
We take the glasses perturbation as an example to elaborate on this step. We begin with 6,000 pairs of face images as original inputs and use \cref{alg:gen} to put on a synthetic pair of glasses to each image to cover no more than 5\%. We try to get around 100 perturbed samples for each input sample, and thus obtained 100 sample pairs for each original sample. In total, we obtain around 600,000 new test cases.
\end{example}

\paragraph{Test Set from Subset Optimisation.}
At the end of the previous step, we obtain a set of potential adversarial examples $\mathbb{T'}_\gamma$ for a given perturbation $d, \epsilon$, where each $\gamma$ denotes a distinct $d, \epsilon$ combination. However, not all of these adversarial examples are effective and the set is too large to be practical. Therefore, we want to find a smaller, optimal subset of $\mathbb{T'}_\gamma$ that serves our purpose. This optimal subset is denoted as $\mathbb{T}_\gamma$, as seen in \cref{eq:testsuite}.

To determine $\mathbb{T}_\gamma$ given $\mathbb{T'}_\gamma$, we solve an optimisation problem with binary variables, a fitness function, constraints, and non-variable parameters. The non-variable parameters include the known robustness (regarding the perturbation scheme $\gamma$) $(r_\gamma(h_1), r_\gamma(h_2) \ldots)$ of a set of face recognition systems, $(h_1, h_2 \ldots)$. The number of binary variables equals the size of $\mathbb{T'}_\gamma$, \emph{i.e.}, each binary component $z_{(\bm{x}, \bm{x'})}$ in the vector $\bm{z}$ represents whether or not to put the corresponding element ${(\bm{x}, \bm{x'})}$ from $\mathbb{T}'_\gamma$ into $\mathbb{T}_\gamma$. The problem is formulated as
\begin{equation}\label{eq:opt}
\begin{aligned}
    &\min_{\bm{z}} \quad \phi_1\Big(\phi_2(\bm{z}, \bm{h}), r_\gamma(\bm{h})\Big) +  \lambda_1~\phi_3(\bm{z}, \bm{h})\\
    &\begin{aligned}
    \mbox{subject~to}\quad &\norm{\bm{z}}_1 \geq k_{\min}, \\
    \quad &\norm{\bm{z}}_1 \leq k_{\max}, \\
    \quad &z_{(\bm{x}, \bm{x'})} \in \set{0, 1}, \forall (\bm{x}, \bm{x'}) \in \mathbb{T}'    
    \end{aligned}\\
    &\begin{aligned}
        \mbox{where } & \bm{z} = [z_{(\bm{x}, \bm{x'})}, \ldots ~~~\forall ~{(\bm{x}, \bm{x'})}\in \mathbb{T}']^T\\
        & \bm{h} = [h_1, h_2, \ldots]^T\\
        & r_\gamma(\bm{h}) = [r_\gamma(h_1), r_\gamma(h_2), \ldots]^T\\
        & \phi_1\big(\bm{a}, \bm{b}\big) = \frac {\sum_{j}(a_j-\bar{a})(b_j-\bar{b})}{\big(\sum_{i'}(a_{i'}-\bar{a})^2 \sum_i (b_i-\bar{b})^2\big)^\frac{1}{2}}\\
        &\phi_2(\bm{z}, \bm{h}) = \left[ \sum_{(\bm{x}, \bm{x'})\in \mathbb{T}'} z_{(\bm{x}, \bm{x'})} \iverson{ h_{\textcolor{magenta}{1}}(\bm{x'}) \neq  y}, \textcolor{magenta}{\ldots}\right]^T\\
        &\phi_3(\bm{z}, \bm{h}) = \operatorname{Mean}_{\bm{h}} \phi_2(\bm{z}, \bm{h}) - \lambda_2 \operatorname{Std}_{\bm{h}} \phi_2(\bm{z}, \bm{h})        
    \end{aligned}
\end{aligned}
\end{equation}
where $\lambda_1, \lambda_2$ are positive constants. The upper and lower bounds of the sum of variables are denoted as $k_{\max}$ and $k_{\min}$, which are used to constrain the number of samples in the final selection. This set selection maximises the Pearson correlation coefficient between the reference robustness and the robustness obtained from evaluating the face recognition systems on the selected set.

To stabilise the optimised selection of examples, we regularise the minimisation process with two more objectives, \emph{i.e.}, (1) to maximise the averaged value, and (2) to maximise the standard deviation across all face recognition systems used for selection. We assign coefficients $\lambda_1$ and $\lambda_2$ for each regularisation.
Regularising the optimisation brings two critical benefits (1) the selected examples are relatively strong adversaries to face recognition systems; (2) and it avoids stopping on a group of examples which can only provide the correct ranking of face recognition systems with very tiny differences. 

Eventually, elements in $\mathbb{T'}_\gamma$ with a positive flag form the selected set $\mathbb{T}_\gamma$, such that
\begin{equation}
    \mathbb{T}_\gamma = \set{{(\bm{x}, \bm{x'})} | {(\bm{x}, \bm{x'})} \in \mathbb{T'}_\gamma, z^*_{(\bm{x}, \bm{x'})} > 0},
\end{equation}
where $\bm{z}^*$ denotes the optimal independent variable vector from Problem \ref{eq:opt}.

Solving the problem using the heuristic optimisation method could result in more than one set that serves as moderately the optimal selection, due to the large number of binary variables. To control the selection of examples, we use a pseudo-random seed to randomly initialise the binary variable (vector) $\bm{z}$. We remark that this prevents malicious users from inflating their robustness score by overfitting to the test suite instance.

The optimisation here helps eliminate the need for a test-selection search in traditional approaches. We remark that our empirical study (see below) shows that the optimal subset, $\mathbb{T}_\gamma$, is \textit{pre}-optimised based on only a few face recognition systems and can be used to evaluate unseen systems. The robustness evaluation using a pre-optimised $\mathbb{T}_\gamma$ is consistent with traditional approaches and significantly speeds up the testing process.

\begin{example}
We continue with the glasses perturbation example. Now that we have 600,000 test cases, we would like to select a small optimal subset among them. We formulate a binary variable problem and eventually, 2,334 cases are shortlisted. Shortlisted test cases are then used to evaluate the robustness of each face recognition system according to \cref{eq:evaluate}. This calculated robustness has over 90\% Pearson correlation with the known robustness of these face recognition systems.
\end{example}

\section{Experiment and Result}
\label{sec:experiment}
Our approach, \textsc{RobFace}, is implemented and evaluated in the following according to the requirements discussed in \cref{sec:preliminary}. We first briefly introduce the setup for generating the set suite and the result $\mathcal{T}$. We then focus on evaluating if \textsc{RobFace} can effectively evaluate the robustness of face recognition systems.

\subsection{Experiment Setup}

\paragraph{Reference Robust Accuracy.}
To evaluate our reported robustness value $\Tilde{r}_\gamma$, we compare it against the reference robust accuracy. The reference robustness accuracy of a system represents the estimation of its robustness. In this study, we use the robust accuracy reported by PGD~\cite{madry2017towards} as the reference. 

We also include CLEVER~\cite{weng2018evaluating} to offer additional perspective on different robustness evaluation approaches. This is a well-known method for verifying robustness through formal Lipschitz constant evaluation. 

\paragraph{Reference approach.}
In the implementation of our method, we use PGD~\cite{madry2017towards} as the reference approach for generating potentially plausible examples and optimising the subset. In the optimisation, the training $r_\gamma$ is the reported robust accuracy of the reference approach. Note that this training $r_\gamma$ could possibly but not necessarily take the value of Reference Robust Accuracy for evaluation.

PGD is generally regarded as one of the most effective and widely applicable white-box adversarial examples search algorithms~\cite{huang2020bridging}. Other existing adversarial attack algorithms such as CW~\cite{carlini2017towards} or algorithms developed in future may also be used as the reference where \cref{alg:gen} can still work similarly.

\begin{table}[t]
\centering
\caption{\label{table:system}
An overlook of face recognition systems studied in this work. We list the name, backbone architecture, head function, and the number of parameters (Prm) of each system.}
\begin{tabular}{llll}
\toprule
\textbf{\textsc{System}}   & \textbf{\textsc{Backbone}} & \textbf{\textsc{Head}} & \textbf{\textsc{\#Prm.}}\\
\midrule
\rnumcap{1} & FaceNet~\cite{schroff2015facenet}   &  Triplet loss~\cite{weinberger2005distance} & 140M \\
\rnumcap{2} & iResNet~\cite{deng2019arcface} &  ArcFace~\cite{deng2019arcface} & 87.14M \\
\rnumcap{3} & EfficientNet~\cite{tan2019efficientnet} & MV-Softmax~\cite{wang2020mis} & 33.44M  \\
\rnumcap{4} & ReXNet~\cite{han2021rethinking} & CurricularFace~\cite{huang2020curricularface}  &  15.20M  \\
\rnumcap{5} & AttentionNet~\cite{wang2017residual} &  AdaCos~\cite{zhang2019adacos} & 98.96M  \\
\rnumcap{6} & GhostNet~\cite{han2020ghostnet} &  MV-Softmax~\cite{wang2018additive} & 26.76M \\
\rnumcap{7} & RepVGG~\cite{ding2021repvgg} & AM-Softmax~\cite{wang2020mis} & 39.94M  \\
\rnumcap{8} & TF-NAS~\cite{hu2020tf}  & MV-Softmax~\cite{wang2020mis} & 39.59M  \\
\rnumcap{9} & LightCNN~\cite{wu2018light} & AM-Softmax~\cite{wang2018additive} & 11.60M  \\
\bottomrule
\end{tabular}
\end{table}
\begin{table}[t]
\centering
\caption{\label{table:dataset}
An overview of standard face verification datasets characteristics in this work.}
\begin{tabular}{llll}
\toprule
\textbf{\textsc{Dataset}}   & \textbf{\textsc{\#Identity}} & \textbf{\textsc{\#Image}} & \textbf{\textsc{\#Pairs}}\\
\midrule
LFW~\cite{huang2008labeled}   & 5,749  & 13,233 & 6,000 \\
CFP-FP~\cite{sengupta2016frontal} & 500 & 7,000 & 7000 \\
AgeDB-30~\cite{moschoglou2017agedb} & 568 & 16,488 & 3000  \\
CPLFW~\cite{zheng2018cross} & 5,749  & 11,652 & 6,000  \\
CALFW~\cite{zheng2018cross} & 5,749  & 12,174 & 6,000  \\
\bottomrule
\end{tabular}
\end{table}
\begin{table}[t]
    \centering
    \caption{Comparison of various perturbation schemes for face images: limitations of existing reference approaches and the general applicability of the proposed method. We remark that although the first reference has applicability to transformations other than $L^p$-norms, necessary extensions are required.}
    \label{tab:accomodation}
    \begin{tabular}{llll}
        \toprule
        \textbf{\textsc{Perturb}} & \textbf{\textsc{Reference-1}} & \textbf{\textsc{Reference-2}} & \textbf{\textsc{RobFace}} \\
        \midrule
        $L^2$-norm &\checkmark & \checkmark & \checkmark \\
        $L^\infty$-norm & \checkmark & \checkmark & \checkmark \\
        Glasses & \checkmark & - & \checkmark \\
        Mask & \checkmark & - & \checkmark \\
        Illumination & \checkmark & - & \checkmark \\
        Radial & \checkmark & - & \checkmark \\
        Age & \checkmark & - & \checkmark \\
        Pose & \checkmark & - & \checkmark \\
        \bottomrule
    \end{tabular}

\end{table}

\pgfplotsset{every tick label/.append style={font=\tiny}}

\begin{figure*}[t]
\centering
\scriptsize
\begin{subfigure}{0.225\textwidth}
\centering
\begin{tikzpicture}

\begin{axis}[
shift,
legend style={at={(1,1)},anchor=north west},
xmin=1, xmax=9,
xtick={1,2,3,4,5,6,7,8,9},xticklabels={I, II, III, IV, V, \underline{VI}, \underline{VII}, \underline{VIII}, IX},
width = \linewidth
]

\addplot+[color=domino,mark=square,]
table[x=system, y= b1mask]
{robustness.dat};


\addplot+[color=catalina_blue,mark=asterisk,]
table[x=system, y= pmask]
{robustness.dat};

\end{axis}
\end{tikzpicture}
\caption{Face mask ($corr =0.99$)}
\end{subfigure}
\begin{subfigure}{0.225\textwidth}
\centering
\begin{tikzpicture}

\begin{axis}[
shift,
legend style={at={(1,1)},anchor=north west},
xmin=1, xmax=9,
xtick={1,2,3,4,5,6,7,8,9},xticklabels={I, II, III, IV, V, VI, \underline{VII}, \underline{VIII}, \underline{IX}},
width = \linewidth
]


\addplot+[color=domino,mark=square,]
table[x=system, y= b1glass]
{robustness.dat};


\addplot+[color=catalina_blue,mark=asterisk,]
table[x=system, y= pglass]
{robustness.dat};

\end{axis}
\end{tikzpicture}
\caption{Glasses ($corr = 0.90$)}
\label{fig:corrglass}
\end{subfigure}
\begin{subfigure}{0.225\textwidth}
\centering
\begin{tikzpicture}

\begin{axis}[
shift,
legend style={at={(1,1)},anchor=north west},
xmin=1, xmax=9,
xtick={1,2,3,4,5,6,7,8,9},xticklabels={I, II, III, IV, V, \underline{VI}, VII, \underline{VIII}, \underline{IX}},
width = \linewidth
]

\addplot+[color=domino,mark=square,]
table[x=system, y= b1linf]
{robustness.dat};

\addplot+[color=mymagenta,mark=triangle,]
table[x=system, y= b2linf]
{robustness.dat};

\addplot+[color=catalina_blue,mark=asterisk,]
table[x=system, y= plinf]
{robustness.dat};

\end{axis}
\end{tikzpicture}
\caption{$L^\infty$ norm ($corr = 0.92$)}
\end{subfigure}
\begin{subfigure}{0.225\textwidth}
\centering
\begin{tikzpicture}

\begin{axis}[
shift,
legend style={at={(1.02,1)},anchor=north west},
xmin=1, xmax=9,
xtick={1,2,3,4,5,6,7,8,9},xticklabels={I, II, III, IV, V, \underline{VI}, \underline{VII}, \underline{VIII}, IX},
width = \linewidth
]

\addplot+[color=domino,mark=square,]
table[x=system, y= b1l2]
{robustness.dat};
\addlegendentry{Reference-1}

\addplot+[color=mymagenta,mark=triangle,]
table[x=system, y= b2l2]
{robustness.dat};
\addlegendentry{Reference-2}

\addplot+[color=catalina_blue,mark=asterisk,]
table[x=system, y= pl2]
{robustness.dat};
\addlegendentry{\textsc{RobFace}}

\end{axis}
\end{tikzpicture}
\caption{$L^2$ norm ($corr = 0.97$)}
\end{subfigure}\\

\begin{subfigure}{0.225\textwidth}
\centering
\begin{tikzpicture}

\begin{axis}[
shift,
legend style={at={(1,1)},anchor=north west},
xmin=1, xmax=9,
xtick={1,2,3,4,5,6,7,8,9},xticklabels={I, II, \underline{III}, \underline{IV}, V, VI, VII, \underline{VIII}, IX},
width = \linewidth
]

\addplot+[color=domino,mark=square,]
table[x=system, y= b1age]
{robustness.dat};


\addplot+[color=catalina_blue,mark=asterisk,]
table[x=system, y= page]
{robustness.dat};

\end{axis}
\end{tikzpicture}
\caption{Age shift ($corr =0.99$)}
\end{subfigure}
\begin{subfigure}{0.225\textwidth}
\centering
\begin{tikzpicture}

\begin{axis}[
shift,
legend style={at={(1,1)},anchor=north west},
xmin=1, xmax=9,
xtick={1,2,3,4,5,6,7,8,9},xticklabels={\underline{I}, II, \underline{III}, IV, V, VI, \underline{VII}, VIII, IX},
width = \linewidth
]

\addplot+[color=domino,mark=square,]
table[x=system, y= b1pose]
{robustness.dat};


\addplot+[color=catalina_blue,mark=asterisk,]
table[x=system, y= ppose]
{robustness.dat};

\end{axis}
\end{tikzpicture}
\caption{Pose shift ($corr =0.99$)}
\end{subfigure}
\begin{subfigure}{0.225\textwidth}
\centering
\begin{tikzpicture}

\begin{axis}[
shift,
legend style={at={(1,1)},anchor=north west},
xmin=1, xmax=9,
xtick={1,2,3,4,5,6,7,8,9},xticklabels={I, II, \underline{III}, \underline{IV}, V, \underline{VI}, VII, VIII, IX},
width = \linewidth
]

\addplot+[color=domino,mark=square,]
table[x=system, y= b1barrel]
{robustness.dat};


\addplot+[color=catalina_blue,mark=asterisk,]
table[x=system, y= pbarrel]
{robustness.dat};

\end{axis}
\end{tikzpicture}
\caption{Radial ($corr =0.97$)}
\end{subfigure}
\begin{subfigure}{0.225\textwidth}
\centering
\begin{tikzpicture}

\begin{axis}[
shift,
legend style={at={(1,1)},anchor=north west},
xmin=1, xmax=9,
xtick={1,2,3,4,5,6,7,8,9},xticklabels={\underline{I}, II, III, IV, V, VI, \underline{VII}, VIII, \underline{IX}},
width = \linewidth
]

\addplot+[color=domino,mark=square,]
table[x=system, y= b1illum]
{robustness.dat};


\addplot+[color=catalina_blue,mark=asterisk,]
table[x=system, y= pillum]
{robustness.dat};

\end{axis}
\end{tikzpicture}
\caption{Illumination ($corr = 0.99$)}
\end{subfigure}
\caption{We show that the proposed approach, \textsc{RobFace}, correlates well with the reference robust accuracies, with Pearson correlation ranging from 0.9 to 0.99. A random group of face systems is used for tuning, and the rest are for testing. Indices of the testing systems are underlined. Reference-1 is the robust accuracy by searching adversarial examples via PGD~\cite{madry2017towards}. Reference-2 is CLEVER~\cite{weng2018evaluating}, a theoretical robustness evaluation approach.
}
\label{fig:correlation}
\end{figure*}
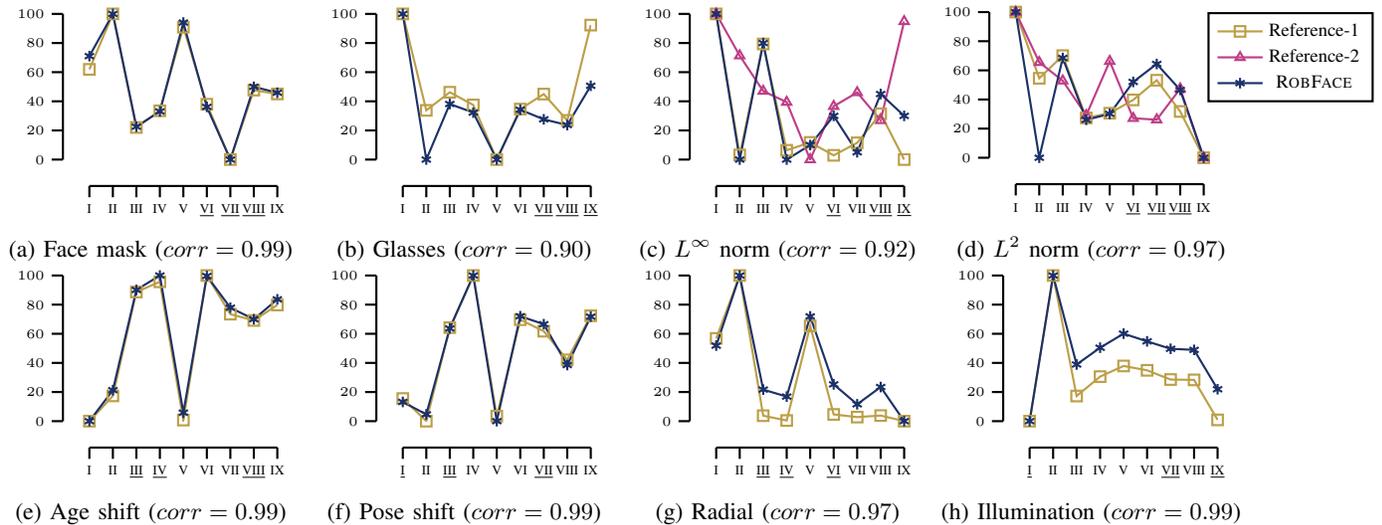






\paragraph{Face recognition systems.} Similar to fitting a machine learning system to data, the proposed method fits the behaviour of a face recognition system. Behaviours of \textit{a small group of} systems are used to tune $\mathcal{T}$, while the others are used to test $\mathcal{T}$. We adopt a list of standard state-of-the-art face recognition systems from the literature~\cite{jain2011handbook}, as shown in \cref{table:system}. The $\mathcal{T}$ optimised for the existing face recognition system can be directly used to test future face recognition systems.

\paragraph{Datasets.}
We adopt a list of popular validation datasets for face recognition to evaluate our approach and the references. In total, these public datasets comprise 28k face pairs, as seen in \cref{table:dataset}. More face pairs with the same format can be used if necessary.

\paragraph{\textsc{RobFace}-01: Perturbation space.}
Following the construction rules of $\mathcal{T}$, we create an instance: \textsc{RobFace}-01. The name is to distinguish it from the name of the proposed overall \emph{approach} to estimating face recognition system robustness. \textsc{RobFace}-01 is a pre-optimised self-contained test suite covering 8 perturbation spaces, as seen in \cref{tab:accomodation}.

\paragraph{\textsc{RobFace}-01: Example Generation.}
\textsc{RobFace-Gen} requires $\mathbb{S},$ $g^\dagger,$ $(d, \epsilon)$ as inputs, and $N, \eta$ as hyper-parameters. Here, $\mathbb{S}$ is a random subset of the union of LFW, CFP-FP, and Age-DB-30. $\mathbb{S}$ contains $\abs{\mathbb{S}}$ pairs of face images to be perturbed, as seen in \cref{alg:gen}. Here, $\abs{\mathbb{S}}=6,000$, covering 37.5\% of the union dataset. We randomly select a face recognition system in \cref{table:system} as $g^\dagger$, and here System-\rnumcap{2} is selected. We also set $N=100$; $\eta$ depends on specific settings of $(d,\epsilon)$. It takes around 460 minutes for an AMD EPYC 7V12 64-Core Processor to generate each $\mathbb{T'}_\gamma$.

\paragraph{\textsc{RobFace}-01: Optimisation.}
The binary variable problem is to select elements in $\mathbb{T'}_\gamma$ to form $\mathbb{T}_\gamma$. The primary objective in Problem \ref{eq:opt} involves calculating the robustness variance across different face recognition systems, as well as the covariance between predicted robustness and known robustness. To this end, the group of face recognition systems to tune the binary variables is randomly sampled from all systems in \cref{table:system}. A hyper-parameter is the size of the tuning group $K_\text{tune} = 5$. The objective is calculated based on a group of tuning systems, and the binary variable is solved with genetic algorithm (GA~\cite{mirjalili2019genetic}, pymoo~\cite{blank2020pymoo}). It takes about $120\pm50$ minutes for an AMD EPYC 7V12 64-Core Processor to go through 1000 iterations in GA to obtain each $\mathbb{T}_\gamma$. Generally, around $1.6\pm0.3$ \% of member of $\mathbb{T'}_\gamma$ is selected as members of $\mathbb{T}_\gamma$, and the optimised test set $\mathbb{T}_\gamma$ for each scheme $(d,\epsilon)$ eventually contains 1,251 to 3,360 face samples ranging from 26,336 samples in total.

\subsection{Research Questions and Answers}
\paragraph{RQ1:}
\textit{How consistent is \textsc{RobFace} with existing approaches on evaluating robustness?}

We use \textsc{RobFace} to evaluate the robustness of each face recognition system in \cref{table:system}, and run reference approaches for these systems. Reference-1 is generally applicable to all perturbation spaces while reference-2 is only applicable to $p$-norm perturbations. In this research question, we consider that we can tune our test suite (\cref{sec:construct}, optimisation \ref{eq:opt}) according to reference. To answer whether \textsc{RobFace} can adequately represent the reference, we first randomly select a tuning group with five systems from the existing face recognition systems (the test group with the remaining systems will be used in RQ2). After that, correlations between \textsc{RobFace} results and reference results are calculated. All perturbation spaces in \cref{tab:accomodation} are illustrated.  We adopt Pearson's correlation coefficient ($corr$) here.

\cref{fig:correlation} illustrates the robustness estimate for different face recognition systems by \textsc{RobFace} as well as references. It shows that the \textsc{RobFace} estimation and empirical estimation correlate well. In general, (for the training group) the correlation between \textsc{RobFace} result and reference-1 ranges from 0.95 to 0.99. Specifically, apart from the perturbation of masks (\cref{fig:correlation}a), age (\ref{fig:correlation}e), and poses (\ref{fig:correlation}f) where a clear match is observed, the correlations regarding the other perturbation spaces are also promising. For instance, regarding the radial perturbation, while the distinction of robustness among systems VII, VIII, IX could be less pronounced than the reference, the ranking of robustness in the reference can be well captured by \textsc{RobFace}. While the match in the illumination perturbation case may seem less obvious due to the gap in between, the correlation is still competitive as a result of correct ranking. Furthermore, the correlation with each reference was higher than the within-reference correlation. Therefore, the decision divergence that the proposed method may induce can be easily tolerated. As mentioned before, reference 2 only works for $L^p$-norms and the correlation is 0.48 and 0.63. Besides, certain systems (e.g., V) could be one of the most robust ones against some perturbation (e.g., pose shift), but the least when it comes to others (e.g., glasses attacks). This aligns with the prior observation~\cite{zhang2023proa}, and could probably be attributed to different feature-extracting architectures.


\begin{summary}[title=RQ1: Goodness of Fit,]
\vspace{-5mm}\textit{\textsc{RobFace} well fits the reference robustness observations with a small discrepancy. }
\end{summary}

\paragraph{RQ2:}
\textit{How well does \textsc{RobFace} generalise to unseen face recognition systems?}

This is a critical question, as \textsc{RobFace} does \emph{not} involve a test-phase search for each face recognition system $g^\dagger$. \textsc{RobFace} uses the pre-optimised sets of samples to estimate the robustness of face recognition systems. Hence, these sets of samples are asked to be strongly transferable, especially to unprecedented systems.

To answer whether our test suite optimised based on a certain tuning group is transferable to unseen systems, we need a random tuning-testing split of existing face recognition systems. As stated in RQ1, a tuning group consists of five systems and a testing group consists of the remaining three systems. Our test suite is only optimised based on the tuning group and then tested on the testing group. In this research question, we measure the correlation between the predicted robustness and known robustness on each group. The result is illustrated in \cref{fig:correlation}. It contains eight plots, and each plot depicts a result regarding a specific perturbation, like glasses perturbation. The indices of testing groups have been underlined, and the correlation under each plot is the overall correlation of all systems. The highest overall correlation is 0.99, observed under face mask perturbation. The lowest overall correlation is 0.90, which is still a significant correlation, observed under glasses perturbation. The group-specific correlations are as follows. The tuning group is $0.99\pm0.04$ and the testing group is $0.95\pm0.05$. We emphasise the overall correlation because the testing-group-only correlation may be inflated when the system rankings are sound within the group but not aligned well with others. For example, the test-group-only correlation under glasses perturbation is 0.99, much higher than its overall correlation, seen in \cref{fig:corrglass}. In contrast, an overall correlation leverages the tuning group as a reference and the robustness predicted for testing systems can be properly evaluated. The well-correlated robustness evaluation of the unseen systems drives us to answer that the proposed method generalises well to unseen face recognition systems.

\begin{figure}[t]
    \centering

    \begin{tikzpicture}
    
    \begin{axis}[
    shift,
    legend style={at={(1.15,1)},anchor=north west, font=\small, cells={align=center}},
    xmin=1, xmax=9,
    xtick={1,2,3,4,5,6,7,8,9},xticklabels={I, II, III, IV, V, VI, \underline{VII}, \underline{VIII}, \underline{IX}},
    width = 0.6\linewidth
    ]
    
    \addplot+[color=domino,mark=square,]
    table[x=system, y= b1glass]
    {robustness.dat};
    \addlegendentry{Reference-1}
    
    
    \addplot+[color=catalina_blue,mark=asterisk,]
    table[x=system, y= pglass]
    {robustness.dat};
    \addlegendentry{\textsc{RobFace}\\\scriptsize $corr=0.90$}
    
    \addplot+[dashed, color=catalina_blue,mark=asterisk,]
    table[x=system, y= pglassrand]
    {robustness.dat};
    \addlegendentry{Random sampling\\\scriptsize $corr=0.19$}
    
    \addplot+[color=catalina_blue!30!white,mark=asterisk,]
    table[x=system, y= pglassnoreg]
    {robustness.dat};
    \addlegendentry{No regularisation\\\scriptsize $corr=0.54$}
    
    \end{axis}
    \end{tikzpicture}

    \caption{Ablation on the proposed methods. We take glasses perturbation as an example. Two ablations are performed. One is to use random sampling to replace \textsc{RobFace-Gen} in the generation phase; the other is to remove regularisation terms in the optimisation phase.}
    \label{fig:ablation}
\end{figure}
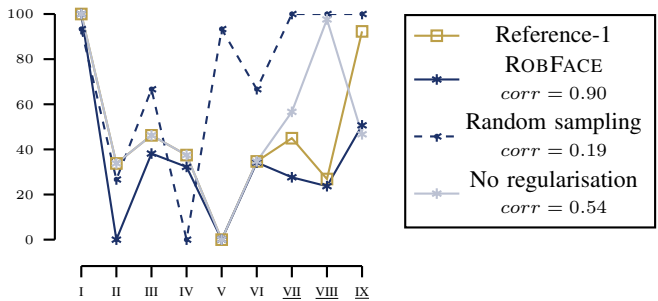

We further conduct an ablation experiment to verify the factors that enable our approach to generalise well. The first is the effectiveness of \textsc{RobFace-Gen}, and the second is regularisation in discrete optimisation. The result of the ablation test is illustrated in \cref{fig:ablation}.
For the former, the sample generation in this work relies on the proposed \textsc{RobFace-Gen} algorithm.  If we replace \textsc{RobFace-Gen} with randomly sampled examples in the glasses perturbation space, the correlation drops from 0.9 to 0.19. Many face systems are assigned with the same level of robustness, such as system I, VII, VIII, IX. The explanation is as follows. The overall advantage of using \textsc{RobFace-Gen} is that all test cases in $\mathbb{T'}_\gamma$ tend to be informative. This is because these test cases cover a range of adversarial losses from low to high. In contrast, other sample generation methods, \emph{e.g.}, neighbourhood random sampling or using standard PGD~\cite{madry2017towards}, may lead to the absence of high-loss examples or low-loss examples. Uneven distribution can lead to indistinguishable results when the generated examples are tested on different systems.

For the latter, regularisation effectively avoids overfitting the tuning group. Using correlation as the only objective may lead to a certain degree of overfitting to the tuning group. As illustrated in \cref{fig:ablation}, if we remove the regularisation in the optimisation, the correlation drops from 0.9 to 0.54. This drop in correlation is a result of overfitted tuning group correlation ($corr=1.0$) and much worse testing group correlation ($corr=-0.83$).  Similar to tuning machine learning systems, our regularisation strategies prevent the parameters from overfitting to the tuning group.

\begin{table*}[t]
\caption{Time taken by different approaches to estimate the robustness of each face system on each perturbation dimension. Average, minimum, and maximum time in seconds is calcuated across each perturbation dimension.}
\label{tab:time}
\centering
\begin{tabular}{llccccccccc}
\toprule
           &      & \multicolumn{9}{c}{\textbf{\textsc{Face system}}}                                                        \\ \cmidrule(lr){3-11}
           &      & I        & II        & III        & IV        & V        & VI        & VII        & VIII        & IX        \\\midrule

\textbf{\textsc{Reference-1}}	&	avg.	&	10569.8	&	3434.8	&	8209.1	&	2406.5	&	13783.9	&	2439.0	&	3763.5	&	6052.7	&	1131.3	\\
	&	min	&	1392.8	&	371.0	&	1009.0	&	202.6	&	459.6	&	193.8	&	633.4	&	515.8	&	64.8	\\
	&	max	&	97370.5	&	14136.7	&	65759.9	&	10880.6	&	81031.7	&	11203.7	&	15753.0	&	31266.5	&	4712.6	\\ \midrule
\textbf{\textsc{Reference-2}}	&	avg.	&	4227.1	&	11072.7	&	8280.9	&	5407.6	&	28006.3	&	5562.9	&	4840.3	&	7139.2	&	6756.8	\\
	&	min	&	4073.5	&	10936.0	&	8154.1	&	5257.6	&	27860.8	&	5373.9	&	4648.7	&	6980.2	&	6568.2	\\
	&	max	&	4380.6	&	11209.3	&	8407.6	&	5557.6	&	28151.8	&	5751.8	&	5031.9	&	7298.3	&	6945.3	\\ \midrule
\textbf{\textsc{RobFace}}	&	avg.	&	24.8	&	41.3	&	30.8	&	18.3	&	167.0	&	18.0	&	14.0	&	25.5	&	22.8	\\
	&	min	&	14.0	&	27.0	&	20.0	&	12.0	&	111.0	&	12.0	&	8.0	&	17.0	&	15.0	\\
	&	max	&	55.0	&	83.0	&	62.0	&	37.0	&	335.0	&	36.0	&	32.0	&	51.0	&	46.0	\\

\bottomrule
\end{tabular}
\end{table*}


\begin{summary}[title=RQ2: Generalisation,]
\vspace{-5mm}
\textit{\textsc{RobFace} generalises well to unseen face recognition systems.}
\end{summary}


\paragraph{RQ3:}
\textit{How efficient is \textsc{RobFace}?}

To answer this question, we compare the efficiency of \textsc{RobFace} versus the two reference methods in terms of time, computing resources, and the required developer's effort.

A prominent advantage of \textsc{RobFace} is that it avoids test-phase searches and uses a pre-optimised test suite to accurately estimate the robustness of various face recognition systems. On average, \textsc{RobFace} requires only 0.3\% of the time used by reference-1, or 0.5\% of the time used by reference-2, as shown in \cref{tab:time}. For perturbations that both \textsc{RobFace} and the reference method can handle, \textsc{RobFace} takes much less time than the reference method, while both use the same CPU (AMD EPYC 7V12 64-core processor) and GPU (A Tesla T4).

Furthermore, we believe that \textsc{RobFace}, as a test suite method, avoids unnecessary trouble in code debugging as much as possible. Using \textsc{RobFace} to estimate the robustness of a system is as straightforward as running it on a general test set. One advantage of the test suite approach is that it is very convenient for system debugging regardless of the machine learning framework, \emph{e.g.}, PyTorch or TensorFlow. In contrast, both reference methods are bond with specific frameworks, adding difficulties in the debugging phase. A common setup for white-box robustness evaluation methods like references is (1) at the local input $x$, the gradients of the neural system are computed by automatic gradient backpropagation; (2) the backpropagated gradients are processed by the white-box evaluator.
Both reference-1 and reference-2 require white-box access to the systems, whereas our approach works in a black-box setting as well.


\begin{summary}[title=RQ3: Efficiency,]
\vspace{-5mm}
\emph{\textsc{RobFace} is around 200 times more efficient than the reference approaches.}
\end{summary}

\paragraph{RQ4:}
\textit{How diverse is the perturbation scheme covered by \textsc{RobFace}?}

\textsc{RobFace} is an approach that accommodates various perturbations, including $p$-norm spaces and other natural transformation spaces. As seen in \cref{tab:accomodation}, \textsc{RobFace} gives a robustness estimation for all listed perturbation scenarios, which matches the number of applicable perturbations for Reference-1, and four times as much as that for Reference-2. \textsc{RobFace} can be easily extended with other kinds of transformations, as long as the transformation is well established.


\begin{summary}[title=RQ4: Perturbation Schemes,]
\vspace{-5mm}
\textit{\textsc{RobFace} has wider applicability than the reference approaches.}
\end{summary}

\subsection{Dealing with Adaptive Attacks}
\begin{figure}[t]
    \centering
    \begin{tikzpicture}
    \begin{axis}[
        ybar,
        bar width=3pt,
        width=0.5\textwidth, 
        height=0.3\textwidth, 
        every outer x axis line/.append style={white}, 
        every outer y axis line/.append style={white}, 
        xlabel={Sets},
        symbolic x coords={Face mask, Glasses, $L^{\infty}$-norm, $L^2$-norm, Age shift, Pose shift, Radial, Illumination},
        xtick=data,
        xtick align=center,
        xtick style={draw=none, thin}, 
        enlarge x limits=0.05, 
        ymin=0,
        ytick distance=20,
        ytick style={draw=none}, 
        axis line style={thin},
        legend style={draw=none, font=\small, at={(0.5,-0.15)}, anchor=north, legend columns=3},
    ]

    \addplot[pattern=north east lines, draw=catalina_blue, fill=catalina_blue!50] coordinates {(Face mask, 94.05) (Glasses, 0.2) ($L^{\infty}$-norm, 10.) ($L^2$-norm, 30.13) (Age shift, 5.839) (Pose shift, 0.5) (Radial, 71.93742) (Illumination, 60.08717497)};
    \addplot[draw=gray, thin, fill=mygray] coordinates {(Face mask, 99.3457) (Glasses, 98.01) ($L^{\infty}$-norm, 98.57) ($L^2$-norm, 97.65) (Age shift, 97.56) (Pose shift, 96.85) (Radial, 97.65) (Illumination, 95)};
    \addplot[draw=catalina_blue] coordinates {(Face mask, 96) (Glasses, 52) ($L^{\infty}$-norm, 35) ($L^2$-norm, 50) (Age shift, 59) (Pose shift, 48) (Radial, 73) (Illumination, 63)};

    \legend{No overfit, With overfit, Change random seed}

    \end{axis}
    \end{tikzpicture}
    \caption{The tested robustness of face recognition system (\emph{e.g.}, V) on \textsc{RobFace}-01 public test suite. We show the tested score before and after the system overfits on the test set. We also test the overfitted system on test sets with a changed seed. Here, the scale is the same as that in \cref{fig:correlation}.}
    \label{fig:adaptive}
\end{figure}
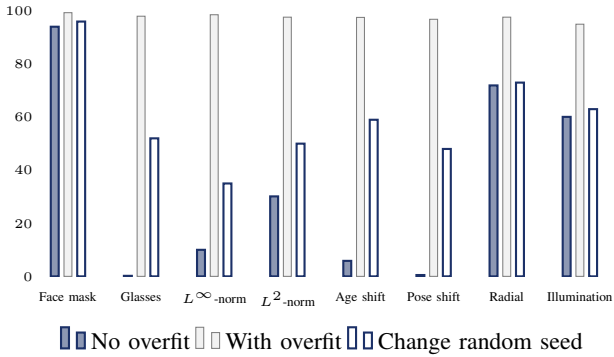
A test suite approach can be subjective to adaptive attacks, \emph{i.e.}, fitting the system directly to the test suite can inevitably help the system to obtain high robustness scores. To counter such an adaptive attack, we can always run the discrete optimisation multiple times with different random seeds to generate a new test set. In fact, with different random seeds, we can create multiple $\mathbb{T}_\gamma$ test suite versions, resulting in a total number of versions calculated by raising the number of versions generated for each perturbation to the power of the number of perturbations used. The randomly provided replicas help avoid overfitting. We conduct an experiment where an existing face recognition system (\emph{e.g.}, V) is fine-tuned on the \textsc{RobFace}-01 public test suite. Expectedly, its ``robustness'' (e.g., to glasses perturbation) is inflated from 0 to 0.98, as shown in \cref{fig:adaptive}. However, once a different random seed is applied, the resultant robustness score drops to 0.52, which has a noticeable gap to the inflated score. The reason why the robustness score is better than the original score of 0 can be explained as an effect of the adversarial training, \emph{i.e.}, by retraining the system on the test suite \textsc{RobFace}-01, the attacker is effectively forced to apply adversarial training and as a result, the system robustness indeed improves. 

\subsection{Threats to Validity}

We use adversarial attacks to measure the true robustness of face recognition systems, which may cause a potential threat to construct validity. Robustness conceptually examines whether there exist adversarial examples in the vicinity of a given input. Given that precisely measuring this quantity is unavailable, adversarial attacks are widely adopted as an approximation measurement. To better infer the relationship or ranking of their true robustness, we conduct adversarial attacks over an extended period, a method shown to effectively bypass various defences~\cite{madry2017towards,croce2020scaling}.

Besides, using datasets to represent the face recognition task may pose threats to external validity. Due to factors like time or region where face images are collected, datasets might have limited generalizability to represent the general distribution in the face recognition task. To compensate for this threat, we have included a series of datasets from multiple sources that diversify the collections to better fit the general distribution.

\section{Related Work}
\label{sec:related}

This work mainly relates to the following works on face recognition and system robustness estimation.

\paragraph{Face recognition systems.}
As a non-intrusive bio-metric technology, face recognition~\cite{wang2021deep} has developed rapidly in recent years, especially in terms of deep neural systems. These deep systems are either trained directly using the triplet loss~\cite{hoffer2015deep}, or trained using the angular marginal loss~\cite{deng2019arcface} for a zero-shot multi-class classification task and then tested on the binary classification task~\cite{liu2019adaptiveface}. Systems are directly trained with triplet loss~\cite{yeung2017improved,trigueros2018enhancing,ge2018deep}. Systems trained with angular marginal loss usually require a feature-extracting backbone that encodes a face image into a vector, and the angular marginal loss tries to separate the vectors for different identities as far as possible~\cite{deng2019arcface,wang2018cosface}. Advanced backbone architectures include resnet~\cite{he2016deep}, iresnet~\cite{deng2019arcface}, transformers~\cite{liu2021swin} etc. Angular marginal loss also has more recent variants like Cos~\cite{wang2018cosface}, sphere~\cite{liu2017sphereface}, or mvsoftmax~\cite{wang2020mis}, aimed at a larger margin among vectors.

\paragraph{Image perturbations.}
Both face recognition studies and broader computer vision research are concerned with image perturbations~\cite{li2019universal}. These perturbations are generally not thought to alter the essence of the image, for example recognising a face shall be invariant to the lighting or pose of the face~\cite{zheng2018cross,li2020light}. This property of being invariant to changes in unimportant factors is often referred to as robustness~\cite{carlini2019evaluating}. A more indicative term would be robustness when there is a difference between benign and malignant predictions~\cite{sanyal2020benign}. Currently, the best-studied image perturbation is $p$-norm noise on the image, as it has the least ambiguity in terms of quantization~\cite{weng2018evaluating}. The study of various other perturbation scenarios besides the $p$-norm has also increased in recent years~\cite{komkov2021advhat,sharif2016accessorize,li2020light,zheng2018cross,sun2022causality,gesi2022code,alrajeh2020combining}. A dataset CIFAR-10C is created to estimate the robustness of image classification systems~\cite{hendrycks2019benchmarking}, but the dataset is fixed and not adaptive to emerging attack algorithms.

\paragraph{System robustness evaluation.}
As increasing attention has been paid to the problem of adversarial perturbation, researchers have begun to study how to quantify the robustness of a system or system. Generally, the robustness is either derived from analysing the adversarial examples of a system in the neighbourhood of test cases~\cite{sharif2016accessorize} or computing some coefficient like Lipschitz on the parameters of the system~\cite{szegedy2013intriguing,hein2017formal,weng2018evaluating}. However, most of these methods are found to be effective for limited perturbation space~\cite{huster2019limitations}, or to be computationally intractable~\cite{virmaux2018lipschitz}. Our work bridges the gap by pre-optimising test cases under each known perturbation, which induces a more comprehensive and efficient robustness estimation. This work is also related to works on analysing neural networks in general~\cite{wang2021robot,asyrofi2021biasfinder,watson2022systematic,usman2022antidotert,tambon2023probabilistic,metzger2022realizing}

\section{Conclusion}
\label{sec:conclusion}
This work is seated in the context of estimating face recognition robustness. We aim to find an efficient evaluation method that works under a large number of perturbations. We propose \textsc{RobFace}, a test-suite approach to estimating the robustness of face recognition systems. The test suite \textsc{RobFace}-01, as proposed in this work, covers more than eight different dimensions of perturbation, including $p$-norm and natural transformations to images. It only takes minutes for \textsc{RobFace} to estimate the robustness of a deep neural face recognition system, while its estimates correlate well with existing approaches. This is because \textsc{RobFace} is the first search-free robustness estimation method where \textsc{RobFace}-01 is pre-optimised with transferable samples to multiple face recognition systems. \textsc{RobFace} is able to accommodate more perturbations for a specific task than existing approaches. We believe it will benefit the agile development of robust face recognition systems in future.

\vfill

\bibliography{custom,favourite,anthology}



\vspace{11pt}

\vspace{-33pt}
\begin{IEEEbiographynophoto}{Ruihan Zhang}
is currently a PhD candidate in the School of Computing and Information Systems at Singapore Management University, supervised by Prof. SUN Jun. Ruihan’s research focuses on evaluating and testing artificial intelligent systems.
\end{IEEEbiographynophoto}

\vspace{-33pt}
\begin{IEEEbiographynophoto}{Jun Sun}
is currently a professor at Singapore Management University (SMU). He received Bachelor and PhD degrees in computing science from National University of Singapore (NUS) in 2002 and 2006. In 2007, he received the prestigious LEE KUAN YEW postdoctoral fellowship. He has been a faculty member since 2010. He was a visiting scholar at MIT from 2011-2012. Jun’s research interests include software engineering, formal methods, program analysis and cyber-security. He is the co-founder of the PAT model checker.
\end{IEEEbiographynophoto}

\end{document}